\documentclass[aps,prx,twocolumn,superscriptaddress,longbibliography]{revtex4-2}

\usepackage{amsmath,amsfonts,bbm}
\usepackage[colorlinks=true,breaklinks=true,linkcolor=magenta,citecolor=magenta,urlcolor=magenta]{hyperref}
\usepackage{comment}
\usepackage{graphicx}
\usepackage{floatrow}
\usepackage{mathtools}
\usepackage{booktabs}
\usepackage{diagbox}
\usepackage{listings}
\usepackage{parskip}
\usepackage{braket}
\usepackage{makecell}
\usepackage{tabularx}
\usepackage{verbatim}
\usepackage{float}
\usepackage{url}
\usepackage{subcaption}
\usepackage{algorithm}
\usepackage{caption}
\usepackage{subcaption}
\usepackage{algpseudocode}
\usepackage[bb=boondox]{mathalfa}
\usepackage{tikz}
\usetikzlibrary{quantikz}

\newcommand{\bts}[1]{{\bf{#1}}}
\newcommand{\crt}[1]{ \hat{a}^\dagger_{#1} }
\newcommand{\dst}[1]{ \hat{a}^{\phantom{\dagger}}_{#1} }
\newcommand{\op}[1]{\hat{#1}}

\begin{document}

\title{Quantum-centric simulation of hydrogen abstraction \\ 
by sample-based quantum diagonalization and entanglement forging}

\author{Tyler Smith}
\address{Integrated Vehicle Systems, Applied Mathematics, Boeing Research \& Technology, Huntington Beach, CA 92647, USA}

\author{Tanvi P. Gujarati}
\address{IBM Quantum, IBM Almaden Research Center, San Jose, CA 95120, USA}

\author{Mario Motta}
\address{IBM Quantum, IBM T. J. Watson Research Center, Yorktown Heights, NY 10598, USA}

\author{Ben Link}
\address{Integrated Vehicle Systems, Applied Mathematics, Boeing Research \& Technology, Huntington Beach, CA 92647, USA}

\author{Ieva Liepuoniute}
\address{IBM Quantum, IBM Almaden Research Center, San Jose, CA 95120, USA}

\author{Triet Friedhoff}
\address{IBM Quantum, IBM T. J. Watson Research Center, Yorktown Heights, NY 10598, USA}

\author{Hiromichi Nishimura}
\address{Integrated Vehicle Systems, Applied Mathematics, Boeing Research \& Technology, Huntington Beach, CA 92647, USA}

\author{Nam Nguyen}
\address{Integrated Vehicle Systems, Applied Mathematics, Boeing Research \& Technology, Huntington Beach, CA 92647, USA}

\author{Kristen S. Williams}
\address{Integrated Vehicle Systems, Applied Mathematics, Boeing Research \& Technology, Huntington Beach, CA 92647, USA}

\author{Javier Robledo Moreno}
\address{IBM Quantum, IBM T. J. Watson Research Center, Yorktown Heights, NY 10598, USA}

\author{Caleb Johnson}
\address{IBM Quantum, IBM T. J. Watson Research Center, Yorktown Heights, NY 10598, USA}

\author{Kevin J. Sung}
\address{IBM Quantum, IBM T. J. Watson Research Center, Yorktown Heights, NY 10598, USA}

\author{Abdullah Ash Saki}
\address{IBM Quantum, IBM T. J. Watson Research Center, Yorktown Heights, NY 10598, USA}

\author{Marna Kagele}
\address{Integrated Vehicle Systems, Applied Mathematics, Boeing Research \& Technology, Huntington Beach, CA 92647, USA}

\begin{abstract}
The simulation of electronic systems is an anticipated application for quantum-centric computers, i.e. heterogeneous architectures where classical and quantum processing units operate in concert. An important application is the computation of radical chain reactions, including those responsible for the photodegradation of composite materials used in aerospace engineering. Here, we compute the activation energy and reaction energy for hydrogen abstraction from 2,2-diphenyldipropane, used as a minimal model for a step in a radical chain reaction. Calculations are performed using a superconducting quantum processor of the IBM Heron family and classical computing resources. To this end, we combine a qubit-reduction technique called entanglement forging (EF) with sample-based quantum diagonalization (SQD), a method that projects the Schr\"{o}dinger equation into a subspace of configurations sampled from a quantum device. In conventional quantum simulations, a qubit represents a spin-orbital. In contrast, EF maps a qubit to a spatial orbital, reducing the required number of qubits by half. We provide a complete derivation and a detailed description of the combined EF and SQD approach, and we assess its accuracy across active spaces of varying sizes upto (39e,39o).
\end{abstract}

\maketitle

\section{Introduction}


Free radicals, because of their high reactivity, can be valuable chemical tools and harmful contaminants. 
Their rich chemical activity stems in part from their tendency to participate in chain reactions.
Such reactions initiate with the formation of radicals from stable species, often through a homolytic bond cleavage triggered by exposure to heat or ultraviolet-visible (UV-vis) light. 
In a subsequent propagation phase, radicals react with stable molecules to form new radicals, e.g. through hydrogen abstraction. 
The reaction ends with the formation of a stable non-radical product from a reaction between radicals.

\begin{figure}[!b]
     \centering
     \includegraphics[width=0.8\columnwidth]{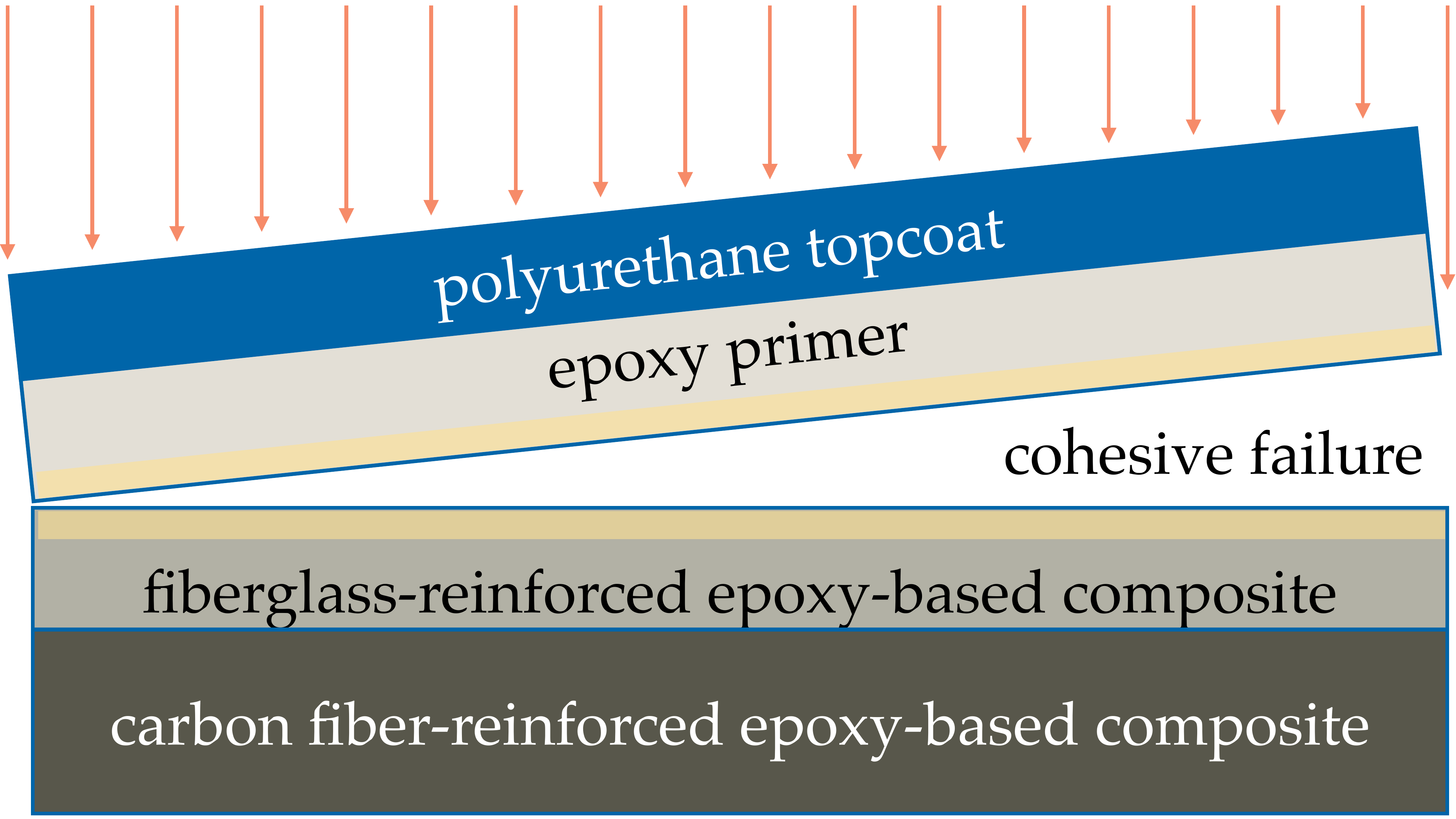}
     \caption{Schematic illustration of the cohesive failure of a composite resin on commercial aircraft due to photo-degradation. While the topcoat and primer layers of paint provide some protection from incident light, after years of exposure enough ultraviolet and visible radiation penetrates through to the composite, ultimately leading to degradation.}
     \label{fig:uv-vis-cartoon}
\end{figure}

Radical chain reactions, and hydrogen abstraction in particular, are major pathways for photo-degradation of composite materials. 
In this context, they lead to phenomena like chain scission, polymer cross-linking and formation of oxidative products, that ultimately cause alterations in mechanical properties, discoloration, brittleness, and loss of performance.
Composite materials used in aerospace-related applications typically consist of a polymer matrix reinforced with carbon or glass fibers, and they are used in the fabrication of aircraft fuselages and wings in lieu of metals due to high strength to weight ratio and corrosion resistance~\cite{mangalgiri1999composite}. Photo-degradation of composite materials employed on in-service aircraft is a substantial issue for airplane manufacturers, leading to hundreds of millions of dollars in maintenance and repair costs. 
For example, the photo-oxidation of the epoxy resin used in aircraft wings results in the peeling of the overlying primer and topcoat paint layers, see Fig.~\ref{fig:uv-vis-cartoon}, causing structural damage and surface blemishes~\cite{mahat2016effects}. 

Studies of this degradation process to date have involved experimental setups to mimic the effects of sunlight damage to composites with paint coatings~\cite{jung2021effect}. Using such setups, research scientists within Boeing Research \& Technologies' Chemical Technologies organization were able to reproduce the damage seen for in-service aircraft in days (while degradation of actual in-service aircraft composites occurs over multiple years)~\cite{gates2003use, rezig2006relationship, schulz2009accelerated}. Experiments leveraging accelerated degradation showed that the addition of a black paint coating between the white top coat and primer offers significant protection from resin degradation and the paint peeling it generates \cite{FAA787}. While this is a convenient palliative measure, a permanent and robust solution will require entirely new material formulations~\cite{parveez2022scientific, zhu2019synthesis}. 
More generally, materials selection for aerospace-related applications is typically performed through testing coupled with engineering judgment~\cite{ arnold2012materials, jayakrishna2018materials}. Qualification of new materials is based on a combination of extensive testing, on-aircraft demonstration or empirical modeling of service life, without necessarily examining the detailed chemistry and physics underlying degradation effects~\cite{miracle2001asm, quilter2001composites}. 


While experimental studies of composite degradation are well-established, ab initio models of epoxy resin degradation have yet to be developed, with current computational models relying on experiment for the determination of many critical parameters ~\cite{lu2018uv, kiil2012model, evans2012statistical}. Few predictive tools are available for simulating material-environment interactions due to the complex nature of such interactions and the range of length scales involved~\cite{meng2023quantum}: aerospace parts are designed at scales ranging from centimeters to several meters, while the degradation mechanisms that lead to part failure in harsh environments -- e.g., corrosion~\cite{gujarati2023quantum}, high-temperature oxidation~\cite{smialek2014oxidation}, ultraviolet degradation~\cite{shi2022analysis} -- occur over only a few nanometers~\cite{DEBORST20081}.
Modeling the mechanisms of polymer degradation from first principles would offer avenues towards predictive characterization for rational, rapid material designs. First-principles simulations may allow the study of reaction pathways for polymer degradation, leading to a deeper understanding of how chain scission, cross-linking, and other forms of degradation unfold~\cite{sai2014first}, and ultimately aid in the design of future resins that incorporate stabilizers or other additives to mitigate the effects of ultraviolet and visible light~\cite{musto2012improving}.

Quantum-centric supercomputing (QCSC)~\cite{alexeev2024quantum} is an innovative computational paradigm in which a quantum computer operates in concert with classical high-performance computing (HPC) resources.
Classical pre-, peri-, and post-processing allows the introduction of quantum subroutines in the economy of classical HPC algorithms, and effectively distributes the computational cost of solution between classical and quantum processors.
The QCSC architecture can significantly extend the domain of applicability of quantum computers through its use in algorithms with carefully designed quantum subroutines, as exemplified by the quantum selected configuration interaction (QSCI)~\cite{kanno2023quantum} and the sample-based quantum diagonalization (SQD) method~\cite{robledo2024chemistry, kaliakin2024accurate, liepuoniute2024quantum, barison2025quantum}.
SQD uses a quantum device to sample electronic configurations. It uses HPC to correct for decoherence-induced breaking of molecular symmetries~\cite{robledo2024chemistry}
as well as to solve the Schr\"{o}dinger equation in a subspace defined by sampled and processed configurations.

In this work, we simulate hydrogen abstraction by a free radical with SQD, and to this end we introduce two methodological extensions.
First, we combine SQD with the entanglement forging method (EF) ~\cite{eddins2022doubling}, a scheme to represent a wavefunction of electrons in $M$ spatial orbitals with $M$-qubit circuits, rather than the $2M$-qubit circuits required by standard calculations.
Second, we generalize the family of quantum circuits used in published literature about EF. Previous works focused on quantum circuits designed primarily for compatibility with noisy quantum devices, resulting in very limited accuracy. 
With improvements in error rates on quantum devices, and the introduction of QCSC algorithms like SQD, we reconsider the structure of the quantum circuits used in EF to achieve a more favorable compromise between hardware compatibility and accuracy.

\section{Methods}

In this Section, we briefly review the SQD and EF methods. We then discuss their combination, and present a more general class of quantum circuits for electronic structure simulations using EF. 

\subsection{Sample-based quantum diagonalization (SQD)}

SQD~\cite{kanno2023quantum, robledo2024chemistry} is a quantum subspace method~\cite{motta2024subspace} that uses a quantum circuit $| \Psi \rangle$ to sample a set of computational basis states $\chi = \{ \bts{x}_1 \dots \bts{x}_d \}$ from the probability distribution 
$p(\bts{x}) = | \langle \bts{x} | \Psi \rangle |^2$ and a classical computer to solve the time-independent Schr\"{o}dinger equation in the subspace spanned by the sampled computational basis states,
\begin{equation}
\label{eq:schrodinger}
\sum_n H_{mn} \, c_n = E c_m \;,\;
H_{mn} = \langle \bts{x}_m | \hat{H} | \bts{x}_n \rangle \;.
\end{equation}
In electronic structure applications, where a computational basis state parametrizes a Slater determinant, the matrix elements $H_{mn}$ can be evaluated efficiently using a classical computer.
Furthermore, because one needs to compute only a few eigenpairs and the matrix $H$ is sparse and diagonally-dominated, Eq.~\eqref{eq:schrodinger} can be solved with Davidson's method.

We remark that 
$(a)$ the spin-resolved particle number operators $\hat{N}_\sigma$ (with $\sigma = \alpha,\beta$ for spin-up/down particles respectively) may not be conserved quantities, i.e. the sampled computational basis states may parametrize Slater determinants with incorrect number of spin-up/down electrons, and
$(b)$ the subspace spanned by the sampled computational basis states may not allow to produce eigenstates of the total spin operator $\hat{S}^2$.
Situation $(a)$ may occur on noiseless quantum devices, if the quantum circuit $| \Psi \rangle$ breaks particle-number symmetries, and on noisy quantum devices, 
where computational basis states are sampled from a probability distribution $\tilde{p}(\bts{x})$ that differs from $p(\bts{x})$.
Similarly, situation $(b)$ may occur on noiseless and noisy quantum devices alike, because Slater determinants are generally not eigenstates of $\hat{S}^2$.

To improve this scenario, SQD employs~\cite{robledo2024chemistry} an iterative self-consistent configuration recovery procedure. Each iteration has two inputs: a set of computational basis states $\chi$ sampled from a quantum device 
and an approximation to the unknown spin-resolved ground-state occupation number distribution $n_{p\sigma} = \langle \Psi_{\mathrm{gs}} | \crt{p\sigma} \dst{p\sigma} | \Psi_{\mathrm{gs}} \rangle$. At each iteration,
\begin{enumerate}
\item for each sampled computational basis state with the wrong spin-resolved particle number, $\bts{x}_k \in \chi$, we compute the deviations $\Delta^k_{p\sigma} = | ( x_k )_{p\sigma} - n_{p\sigma} |$ between the entries of $\bts{x}_k$ and those of the occupation number distribution, 
and use them to define a probability distribution over the set of spin-orbitals $p\sigma$
\item we randomly flip a fixed number of 1s or 0s defined by the difference of the hamming weight of $\bts{x}_k$ and desired number of electrons, according to the probability distribution in point 1, until the spin-resolved particle numbers assume target values, thereby producing a new set of recovered configurations $\chi_{\mathrm{R}}$
\item we sample $K$ subsets (batches) from $\chi_{\mathrm{R}}$, that we call $\chi_b$ with $b=1 \dots K$.
 Each batch yields a subspace $\mathcal{S}_b$ of dimension $d$, in which we project the Hamiltonian as in Eq.~\eqref{eq:schrodinger}, producing an approximation $E_b, | \psi_b \rangle$ to the ground eigenpair.
\item we use the lowest energy across batches, $\mathrm{min}_b \, E_b$, as the best approximation to the ground-state energy, and we use the states $| \psi_b \rangle$ to update the occupation number distribution,
\begin{equation}
n_{p\sigma} = \frac{1}{K} \sum_{b=1}^K \langle \psi_b | \crt{p\sigma} \dst{p\sigma} | \psi_b \rangle
\;.
\end{equation}
\item we repeat points 1-4 until convergence of the energy $\mathrm{min}_b \, E_b$ and the occupation number distribution or for a fixed number of configuration recovery iterations.
\end{enumerate}
In the first iteration of configuration recovery, instead of points 1-2, we postselect~\cite{huggins2021efficient} computational basis states in $\chi$ based on particle number. In addition to the above mentioned configuration recovery steps, we also include the carryover method to improve convergence in energies ~\cite{shirakawa2025closedloop}. To accommodate for long computational times taken for diagonalization with limited compute memory, we relaxed step 4 above and treat each batch independently to allow for independent updates of the occupation number distribution instead of averaging over all batches. In the current formulation of SQD, total spin conservation can only be approximately enforced by adding a penalty term to the Schr\"{o}dinger equation, that penalizes ground-state approximations with expectation value of $\hat{S}^2$ deviating from a target value.

\subsection{Entanglement forging (EF)}

EF~\cite{bravyi2016trading,eddins2022doubling} is an algorithm to represent a $2M$-qubit wavefunction as multiple $M$-qubit wavefunctions embedded in a classical computation.
It is an example of a broader class of algorithms, including quantum embedding~\cite{sun2016quantum}, that partition a system into a collection of weakly interacting clusters and correlate the results of each cluster to reconstruct properties of the original system. 
EF considers wavefunctions of the form
\begin{equation}
\label{eq:forging}
| \Psi \rangle = \sum_\mu c_\mu \ket{u_\mu} \otimes \ket{v_\mu} \;.
\end{equation}
In the original formulation, Eq.~\eqref{eq:forging} is derived from a singular-value decomposition of a wavefunction, implying that the vectors $u_\mu$ and $v_\mu$ are orthonormal and the coefficients $c_\mu$ real and positive.
For reasons of future convenience, we will consider non-orthogonal vectors and complex coefficients.

Eq.~\eqref{eq:forging} allows to write expectation values of $2M$-qubit operators starting from expectation values of $M$-qubit operators, e.g.
\begin{equation}
\label{eq:ef_average}
\begin{split}
\langle \Psi | \hat{A} \otimes \hat{B} | \Psi \rangle 
&= \sum_\mu |c_\mu|^2 a_\mu b_\mu 
+ \sum_{\substack{pr=0 \\ \mu \neq \nu}}^3 c^*_\mu c_\nu \gamma_{\mu\nu}^p \delta_{\mu\nu}^r a_{\mu\nu}^p b_{\mu\nu}^r
\end{split}
\end{equation}
involving the following terms
\begin{equation}
\begin{split}
a_\mu &= \langle u_\mu | \hat{A} | u_\mu \rangle \;,\; a_{\mu\nu}^p = \langle u_{\mu\nu}^p | \hat{A} | u_{\mu\nu}^p \rangle \;, \\
b_\mu &= \langle v_\mu | \hat{B} | v_\mu \rangle \;,\; b_{\mu\nu}^r = \langle v_{\mu\nu}^r | \hat{B} | v_{\mu\nu}^r \rangle \;, \\
\end{split}
\end{equation}
and the following normalized $M$-qubit states,
\begin{equation}
\label{eq:superpositions}
\begin{split}
\ket{u_{\mu\nu}^p} &= \frac{ \ket{u_\mu} + i^p \ket{u_\nu} }{U_{\mu\nu}^p} \;, \\
U_{\mu\nu}^p &= \sqrt{2 + 2 \, \mbox{Re}[ i^p \langle u_\mu | u_\nu \rangle ] } \;, \\
\gamma_{\mu\nu}^p &= \frac{(-i)^p (U_{\mu\nu}^p)^2}{4} \;, \\
\end{split}
\end{equation}
and
\begin{equation}
\begin{split}
\ket{v_{\mu\nu}^p} &= \frac{ \ket{v_\mu} + i^p \ket{v_\nu} }{V_{\mu\nu}^p} \;, \\
V_{\mu\nu}^p &= \sqrt{2 + 2 \, \mbox{Re}[ i^p \langle v_\mu | v_\nu \rangle ] } \;, \\
\delta_{\mu\nu}^p &= \frac{(-i)^p (V_{\mu\nu}^p)^2}{4} \;. \\
\end{split}
\end{equation}
Further simplifications, occurring e.g. if the coefficients $c_\mu$ are real and positive, are detailed in Ref.~\cite{eddins2022doubling}.

\subsection{Combination between EF and SQD}
\label{sec:combination}

In published literature, EF was used to compute expectation values while the central functionality of SQD is sampling from probability distributions.
To combine SQD with EF, we introduce the projectors $| \bts{x} \rangle \langle \bts{x}|$ and $| \bts{y} \rangle \langle \bts{y}|$ acting on the two partitions in Eq.~\eqref{eq:ef_average} to write the probability distribution 
$p(\bts{x}, \bts{y}) = |\langle \bts{x}, \bts{y} | \Psi \rangle|^2$ as
\begin{equation}
\label{eq:ef_pdf}
\begin{split}
p(\bts{x}, \bts{y})
&= \sum_\mu |c_\mu|^2 p(\bts{x}|u_\mu) \, p(\bts{y} | v_\mu) \\
&+ \sum_{\substack{pr=0 \\ \mu \neq \nu}}^3 c^*_\mu c_\nu \gamma_{\mu\nu}^p \delta_{\mu\nu}^r p(\bts{x} | u_{\mu\nu}^p) p(\bts{y} | v_{\mu\nu}^r ) \\
&= \sum_I r_I \, p_I( \bts{x} ) \, q_I( \bts{y} )
\end{split}
\end{equation}
where $p(\bts{z} | \psi) = | \langle \bts{z} | \psi \rangle |^2$ is the probability to measure a length-$M$ bitstring $\bts{z}$ conditional to the preparation of an $M$-qubit
register in the state $|\psi\rangle$. In the last line, we introduce the compact notation $\sum_I r_I \, p_I( \bts{x} ) \, q_I( \bts{y} )$ to avoid clutter.
Eq.~\eqref{eq:ef_pdf} is not a compound probability distribution, because the coefficients $q_I$ are not, in general, a probability distribution. 
However, Eq.~\eqref{eq:ef_pdf} may be approximated with a compound probability distribution of the form
\begin{equation}
\label{eq:ef_pdf_compound}
\begin{split}
\tilde{p}(\bts{x}, \bts{y}) = \sum_I P_I \, p_I( \bts{x} ) \, q_I( \bts{y} )
\end{split}
\end{equation}
where the coefficients $\{ P_I \}_I$ are the entries of a probability distribution.
Finding such probability distribution requires minimizing the log-likelihood function $\sum_{\bts{x} \bts{y}} \left[ \, p(\bts{x}, \bts{y}) - \tilde{p}(\bts{x}, \bts{y}) \, \right]^2$~\cite{smolin2012efficient} ,
which in turn requires knowledge of the inner products between the conditional probability distributions in Eq.~\eqref{eq:ef_pdf}.
This information cannot be efficiently obtained by sampling, as it requires post-selection over measurement outcomes. Instead, in this study, we elected to use the approximation
\begin{equation}
P_I = \frac{ \max\big[ 0, \mbox{Re} \left( r_I \right) \big] }{ \sum_J \max\big[ 0, \mbox{Re} \left( r_J \right) \big] } \;.
\end{equation}
This choice approximates the process of drawing samples from Eq.~\eqref{eq:ef_pdf} through the process of drawing samples from Eq.~\eqref{eq:ef_pdf_compound}.
More specifically, fixing a total number $N$ of shots, we can $(i)$ sample the distribution $P_I$ for $N$ times and record the numbers $N_I$ of times each of the $I$'s has been sampled, and 
$(ii)$ sample each distribution $p_I( \bts{x} ) \, q_I( \bts{y} )$ for $N_I$ times.

\subsection{A more general family of EF wavefunctions}


EF was used to simulate ground-state potential energy curves of $\mathrm{H_2O}$~\cite{eddins2022doubling}, the activation energy of a Diels-Alder reaction~\cite{liepuoniute2024simulation}, 
and the ground- and excited-state properties of $\mathrm{H_3S^+}$~\cite{bauer2020quantum} and substituted aromatic heterocycles~\cite{castellanos2024quantum}, in the framework of the 
variational quantum eigensolver (VQE) and quantum subspace expansion (QSE) based on single and double electronic excitation operators.
In these studies, the states $u_\mu$ and $v_\mu$ in Eq.~\eqref{eq:forging} were of the form
\begin{equation}
\label{eq:forging_hop_bit}
| u_\mu \rangle = | v_\mu \rangle = \prod_{(pr) \in S} v_{pr}(\theta_{pr}) | \bts{x}_\mu \rangle
\;,
\end{equation}
i.e. they resulted from the action, over a ``bitstring state'' $| \bts{x}_\mu \rangle$, of two-qubit gates called ``hopgates'' -- here denoted $v(\theta)$ and acting on pairs $(pr)$ of qubits in a set $S$.
Such a choice, primarily dictated by compatibility with quantum device connectivity and coherence, resulted in low-accuracy simulations akin to multi-configurational self-consistent field (MCSCF) with 2-3 configurations, 
and prompted the use of expensive post-processing operations to increase accuracy.
In this work, we consider a more general class of quantum circuits for EF calculations, beyond the combination of bitstring states and hopgates employed in published literature.
Such circuits can be used in conventional EF calculations or in combination with SQD.

The starting point of this generalization is the observation that a suitable manipulation of a unitary coupled cluster doubles (uCCD) wavefunction yields a wavefunction that is very reminiscent of the EF state Eq.~\eqref{eq:forging}, 
albeit with clear differences in the structure of the wavefunctions $\ket{u_\mu}$ and $\ket{v_\mu}$, suggesting a natural and compelling generalization of Eq.~\eqref{eq:forging_hop_bit}.
To this end, let us consider the uCCD wavefunction
\begin{equation}
| \Psi_{\mathrm{uCCD}} \rangle = e^{\op{T}-\op{T}^\dagger} | \bts{x}_{\mathrm{HF}} \rangle
\;,
\end{equation}
where $| \bts{x}_{\mathrm{HF}} \rangle$ denotes the Hartree-Fock state (labeled by a bitstring in the basis of molecular orbitals), and
\begin{equation}
\label{eq:t_op}
\begin{split}
\op{T}
&= (t_{2\alpha\alpha})_{aibj} \op{E}^{\alpha\alpha}_{aibj} + (t_{2\beta\beta})_{AIBJ} \op{E}^{\beta\beta}_{AIBJ} \\
&+ (t_{2\alpha\beta})_{aiBJ} \op{E}^{\alpha\beta}_{aiBJ} \\
&= \op{T}_{\alpha\alpha} + \op{T}_{\beta\beta} + \op{T}_{\alpha\beta} \;.
\end{split}
\end{equation}
In the previous equation we used Einstein's summation convention, labeled spin-up/down occupied orbitals with $ij$/$IJ$ and spin-up/down virtual orbitals with $ab$/$AB$,
and denoted excitation operators as 
\begin{equation}
\begin{split}
\op{E}^\sigma_{prqs} &= \crt{p\sigma} \crt{q\sigma} \dst{s\sigma} \dst{r\sigma} \;,\; \op{E}^{\alpha\beta}_{aiBJ} = \crt{a\alpha} \crt{B\beta} \dst{J\beta} \dst{i\alpha} \;.
\end{split}
\end{equation}
With a simple manipulation (derived in Appendix~\ref{sec:appendix_gef}) one can write 
\begin{equation}
\label{eq:sum_of_squares}
\op{T} - \op{T}^\dagger = (\op{T}_{\alpha\alpha} - \op{T}_{\alpha\alpha}^\dagger) + (\op{T}_{\beta\beta} - \op{T}_{\beta\beta}^\dagger) + \sum_\delta \frac{\op{X}_\delta^2}{2}
\end{equation}
where $\op{X}_\delta$ is a spin-unrestricted one-body operator. Employing a primitive Trotter approximation~\cite{trotter1959product} and a Hubbard-Stratonovich transformation~\cite{hubbard1959calculation,stratonovich1957method},
\begin{equation}
e^{\frac{\op{X}^2_\delta}{2}} = \int_{-\infty}^{\infty} dy_\delta \, \frac{e^{- \frac{y_\delta^2}{2}}}{\sqrt{2\pi}} \, e^{y_\delta \op{X}_\delta}
\;,\;
\end{equation}
one can approximate the CCD wavefunction as
\begin{equation}
\label{eq:uCCSD_forging}
| \Psi_{\mathrm{uCCD}} \rangle = \int d{\bf{y}} \, p({\bf{y}}) \, \left( e^{\op{T}_{\alpha\alpha} - \op{T}^\dagger_{\alpha\alpha}} \otimes e^{\op{T}_{\beta\beta} - \op{T}^\dagger_{\beta\beta}} \right) \, | \Phi(\bts{y}) \rangle
\end{equation}
where $| \Phi(\bts{y}) \rangle = e^{ \sum_\delta y_\delta \hat{X}_\delta} | \bts{x}_{\mathrm{HF}} \rangle$ is an unrestricted Slater determinant (see Appendix~\ref{sec:appendix_gef} for a derivation).

Eq.~\eqref{eq:uCCSD_forging} is reminiscent of Eq.~\eqref{eq:forging} and Eq.~\eqref{eq:forging_hop_bit}, with 
$(a)$ an integral over continuous variables instead of a discrete summation, 
$(b)$ a tensor product of unitary CCD operators instead of hopgates, and 
$(c)$ non-orthogonal Slater determinants replacing ``bitstring states''.
It is also reminiscent of the representation of the imaginary-time evolution operator used in 
auxiliary-field quantum Monte Carlo~\cite{zhang2003quantum,motta2018ab} calculations,
although in that context the same-spin terms of the Hamiltonian are also subjected to a Hubbard-Stratonovich transformation.

Eq.~\eqref{eq:uCCSD_forging} suggests a generalization of Eq.~\eqref{eq:forging_hop_bit}, 
where the states $| u_\mu \rangle$ and $| v_\mu \rangle$ have the form
\begin{equation}
\label{eq:gef_state}
\begin{split}
| u_\mu \rangle &= e^{\op{T}_{\alpha\alpha} - \op{T}^\dagger_{\alpha\alpha}} e^{ \op{X}^\mu_\alpha }  | \bts{x}_{\mathrm{HF,\alpha}} \rangle = e^{\op{T}_{\alpha\alpha} - \op{T}^\dagger_{\alpha\alpha}}  | d_{\mu \alpha} \rangle \;, \\
| v_\mu \rangle &= e^{\op{T}_{\beta\beta} - \op{T}^\dagger_{\beta\beta}}     e^{ \op{X}^\mu_\beta }  | \bts{x}_{\mathrm{HF,\beta}} \rangle = e^{\op{T}_{\beta\beta} - \op{T}^\dagger_{\beta\beta}} | d_{\mu \beta} \rangle \;. \\
\end{split}
\end{equation}
In this equation, $\op{X}^\mu_\sigma = \sum_{pr} X^{\mu \sigma}_{pr} \crt{p\sigma} \dst{r\sigma}$ 
is a one-body operator acting on spin-$\sigma$ spin-orbitals, so the states 
$| d_{\mu \alpha} \rangle$ and $| d_{\mu \beta} \rangle$ are Slater determinants~\cite{thouless1960stability,balian1969nonunitary}.

\begin{figure*}[t!]
    \centering
    \includegraphics[width=0.9\linewidth]{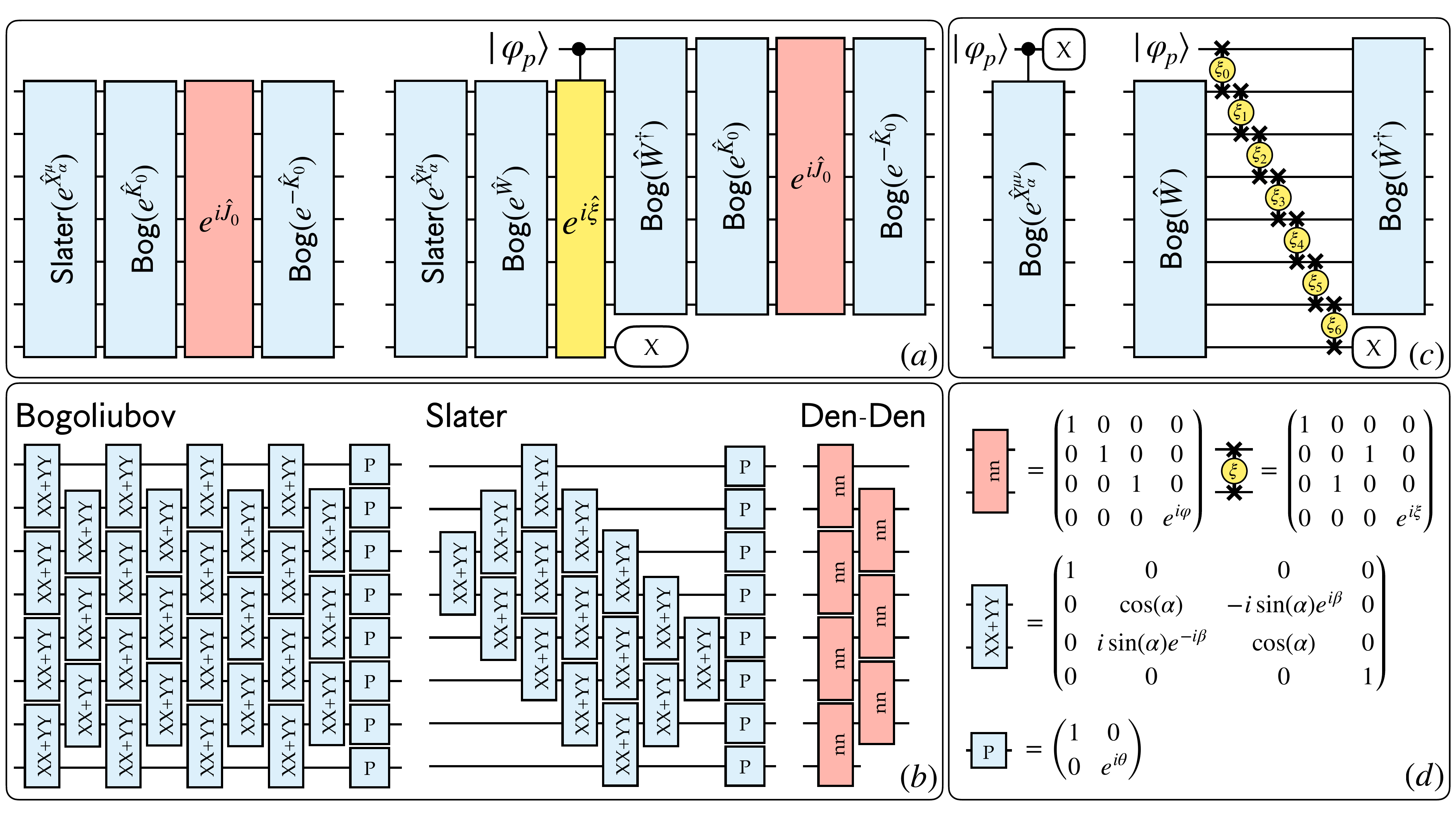}
    \caption{(a) Quantum circuits to prepare the states $u_\mu$ (left) and $u_{\mu\nu}^p$ (right), with $| \varphi_p \rangle = \frac{ | 0 \rangle + i^p | 1 \rangle }{ \sqrt{2} }$. (b) Compilation of an orbital rotation (left) and a quantum circuit to prepare a Slater determinant (middle) into XX+YY and phase gates (blue symbols) and of a density-density interaction (abbreviated ``Den-Den'') into number-number gates with linear qubit connectivity (right, red symbols). (c) quantum circuit to apply a controlled orbital rotation (left) and its compilation into orbital rotations and the controlled exponential of a diagonal one-body operator (right). (d) Definition of one- and two-qubit gates in panels (a)-(c).}
    \label{fig:superposition_circuits}
\end{figure*}

When $\op{T}_{\alpha\alpha} = \op{T}_{\beta\beta} = 0$, the wavefunction Eq.~\eqref{eq:gef_state} takes the form
$\sum_\mu c_\mu | d_{\mu \alpha} \rangle \otimes | d_{\mu \beta} \rangle$ and reduces to resonating Hartree-Fock (ResHF)~\cite{bremond1964hartree,fukutome1988theory} or the non-orthogonal configuration interaction (NOCI)~\cite{malmqvist1986calculation,thom2009hartree,sundstrom2014non} provided that the Slater determinants and coefficients are variationally optimized.
Motivated by the relationship with ResHF and coupled-cluster theory, in this work we
$(i)$ define $| d_{\mu \alpha} \rangle \otimes | d_{\mu \beta} \rangle$ and $c_\mu$ using the output of a ResHF calculation, and
$(ii)$ define $\op{T}_{\alpha\alpha}$ and $\op{T}_{\beta\beta}$ using the output of a classical CCSD calculation (see Appendix~\ref{sec:appendix_gef} for additional details).


\subsubsection{Quantum circuits}

Implementing the EF wavefunction in Eq.~\eqref{eq:gef_state} requires constructing quantum circuits to $(i)$ prepare a Slater determinant, e.g. $| d_{\mu \alpha} \rangle$, or a superposition of Slater determinants, e.g. 
\begin{equation}
\label{eq:superpos}
| d_{\mu\nu \alpha}^p \rangle = \frac{| d_{\mu \alpha} \rangle + i^p | d_{\nu \alpha} \rangle}{U_{\mu\nu}^p} \;,
\end{equation}
with $U_{\mu\nu}^p$ as in Eq.~\eqref{eq:superpositions} and $(ii)$ apply a uCCD operator.
These circuits are shown in Fig.~\ref{fig:superposition_circuits}$a$, and described in detail in this Section.

\paragraph{Slater determinants.} It is well known~\cite{reck1994experimental,clements2016optimal,jiang2018quantum,ffsim}
that, in the standard Jordan-Wigner representation, the exponential of a one-body operator, e.g. $\op{X}^\mu_\alpha$ is implemented by a ``Bogoliubov'' quantum circuit.
This circuit is exemplified in Fig.~\ref{fig:superposition_circuits}$b$ and labeled $\mathsf{Bog}(\op{X}^\mu_\alpha)$. The Slater determinant $| d_{\mu \alpha} \rangle$ can be prepared by a more economical circuit~\cite{reck1994experimental,clements2016optimal,jiang2018quantum,ffsim} whose action on $| \bts{x}_{\mathrm{HF,\alpha}} \rangle$ coincides with that of $\mathsf{Bog}(\op{X}^\mu_\alpha)$, exemplified in Fig.~\ref{fig:superposition_circuits}$b$ and labeled $\mathsf{Slater}(\op{X}^\mu_\alpha)$.

\paragraph{Superpositions of Slater determinants.} Preparing a state of the form Eq.~\eqref{eq:superpos} is more subtle. First, let us write 
\begin{equation}
\label{eq:superposition_of_dets}
| d_{\mu\nu \alpha}^p \rangle = \frac{ I + i^p e^{\op{X}^{\mu\nu}_\alpha} }{ U_{\mu\nu}^p } | d_{\mu \alpha} \rangle
\;,
\end{equation}
where $e^{\op{X}^{\mu\nu}_\alpha} = e^{\op{X}^{\nu}_\alpha} e^{-\op{X}^{\mu}_\alpha}$ is an orbital rotation. To apply the linear combination of unitaries $I + i^p e^{\op{X}^{\mu\nu}_\alpha}$ to the state $| d_{\mu \alpha} \rangle$, one can use the quantum circuit in Fig.~\ref{fig:superposition_circuits}$c$ (upper panel, see Appendix~\ref{sec:appendix_gef} for a proof). To implement the controlled orbital rotation, we recall that $\op{X}^{\mu\nu}_\alpha$ is an anti-Hermitian operator and thus can be diagonalized by an orbital rotation $\op{W}$, leading to
\begin{equation}
\label{eq:super_circuit}
\mathsf{c} \left[ e^{\op{X}^{\mu\nu}_\alpha} \right]
= 
\op{W}^\dagger \mathsf{c} \left[ e^{i \sum_q \xi_q \crt{q \alpha} \dst{q \alpha} } \right]\op{W} 
\;,
\end{equation}
where $\xi_q$ are the eigenvalues of $\op{X}^{\mu\nu}_\alpha$ and $\mathsf{c}[\op{U}] = | 0 \rangle \langle 0 | \otimes I + | 1 \rangle \langle 1 | \otimes \op{U}$ denotes the controlled version of a quantum circuit. To implement the right-hand side of Eq.~\eqref{eq:super_circuit}, we use the circuit shown in Fig.~\ref{fig:superposition_circuits}$c$ (lower panel, see Appendix~\ref{sec:appendix_gef} for additional details).

\paragraph{Unitary coupled-cluster.} Let us consider the same-spin CCD operators $\exp(\op{T}_{\sigma\sigma} - \op{T}^\dagger_{\sigma\sigma})$. Through a low-rank decomposition of the $t_2$ amplitudes~\cite{motta2021low} and a Trotter approximation, these unitary operators can be written as
\begin{equation}
\label{eq:CCD_low_rank}
e^{\op{T}_{\sigma\sigma} - \op{T}^\dagger_{\sigma\sigma}} \simeq \prod_{\mu=1}^L e^{\hat K_\mu} e^{i\hat J_\mu} e^{-\hat K_\mu}
\end{equation}
where the operators
\begin{equation}
\label{eq:ucj}
\hat K_\mu = \sum_{pr} \kappa^\mu_{pr}(t_2) \, \crt{p\sigma} \dst{r\sigma}
\;\; , \;\;
\hat J_\mu = \sum_{pr} J^\mu_{pr}(t_2) \, \hat{n}_{p\sigma} \hat{n}_{r\sigma}
\end{equation}
are determined by the $t_2$ amplitudes. The operators in Eq.~\eqref{eq:ucj} are, respectively, an orbital rotation and a ``density-density interaction''. The latter can be implemented by a circuit of $\mathsf{ZZ}$ rotations with all-to-all connectivity.
Eq.~\eqref{eq:CCD_low_rank} motivated the unitary cluster Jastrow ansatz~\cite{matsuzawa2020jastrow}, in which the number of terms in the product can be chosen by the user rather than resulting from the $t_2$ amplitudes and the number of steps in the Trotter approximation, and the local unitary cluster Jastrow (LUCJ) ansatz~\cite{motta2023bridging}, in which the density-density operators $\hat J_\mu$ requiring all-to-all qubit connectivity or costly SWAP networks~\cite{o2019generalized} are truncated by replacing the dense matrix $J^\mu_{pr}$ with a sparse matrix. While, in near-term implementation, the only elements of $J^\mu_{pr}$ are those that may be implemented without any SWAP networks (see Fig.~\ref{fig:superposition_circuits}$b$ for an example), as error rates on quantum devices decrease, one may consider a hierarchy of increasingly dense matrices converging towards the UCJ limit (e.g. banded matrices). The quantum circuits that implement the LUCJ ansatz with $L=1$ are shown in Fig.~\ref{fig:superposition_circuits}$a$.

\begin{table*}
\begin{tabular}{ccccc}
\hline\hline
           & qubits & XX+YY & nn & depth \\
\hline
LUCJ       & $2M$   & $\sum_\sigma N_\sigma (M-N_\sigma) + L \left\{ 2 M (M-1) \right\}$ & $L \left\{ 2 (M-1) + \frac{M}{4} \right\}$ & $ (M-1) + L \left\{ M+3 \right\}$ \\
EF (ind)   &  $M$   & $N_\sigma (M-N_\sigma) + L \left\{ M (M-1) \right\}$               & $L \left\{ 2 (M-1) + \frac{M}{4} \right\}$ & $ (M-1) + L \left\{ M+2 \right\}$ \\
EF (super) &  $M$   & $N_\sigma (M-N_\sigma) + (M+1) (M-1) + L \left\{ M (M-1) \right\}$ & $L \left\{ (M-1) \right\}$                 & $3(M-1) + L \left\{ M+2 \right\}$ \\
\hline\hline
\end{tabular}
\caption{Quantum resources to execute a standard LUCJ circuit on a lattice with heavy-hex qubit connectivity (first row) and to prepare the individual states $u_\mu$, $v_\mu$ and the superposition states $u_{\mu\nu}^p$, $v_{\mu\nu}^r$ in EF (second and third row, labeled ``ind'' and ``super'' respectively). $M$, $N_\sigma$, and $L$ denote the number of spin-$\sigma$ electrons, spatial orbitals, and layers of the LUCJ circuit, respectively.}
\label{table:resources}
\end{table*}

\paragraph{Quantum resource estimate}

Quantum resources (number of qubits, number of 2-qubit XX+YY and nn gates, and depth of the 2-qubit gate) to execute standard LUCJ and EF circuits are listed in Table~\ref{table:resources}.
Circuits for preparing individual states $u_\mu$, $v_\mu$ have roughly half the gates of standard LUCJ circuits and a marginally lower depth.
Circuits for preparing superposition states $u_{\mu\nu}^p$, $v_{\mu\nu}^r$ have roughly half the gates of standard LUCJ circuits plus $(M+1)(M-1)$ additional gates and higher depth, specifically $2 (M-1)$ more layers of two-qubit gates. Since the $(M+1)(M-1)$ additional gates roughly correspond to the cost of a Bogoliubov circuit, the standard LUCJ circuit has more gates for $L \geq 2$. The decreased qubit count, compared to standard LUCJ, gives more freedom in the choice of the qubit layout and protection against cross-talk errors.

\subsection{Target reaction}


\begin{figure*}
    \centering
    \includegraphics[width=0.75\linewidth]{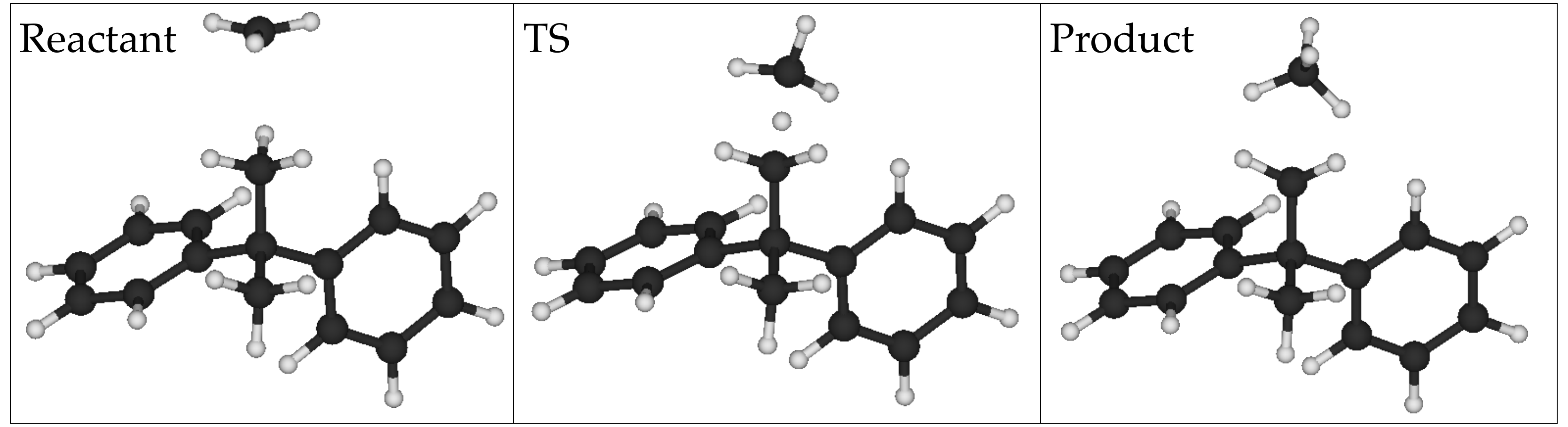}
    \caption{Left to right: reactant, transition state, and product geometries for the hydrogen abstraction reaction considered in this study.}
    \label{fig:radical-epoxy-model}
\end{figure*}

This study is motivated by the photo-oxidation of the epoxy resin DGEBA~\cite{rivaton1997photo2}. This process is a multistep chain reaction~\cite{yousif2013photodegradation}, described in detail in Appendix~\ref{sec:dgeba},
comprising a propagation phase in which a radical from the environment abstracts a hydrogen from a primary methyl group on the DGEBA subunit.
Rivaton et al~\cite{rivaton1997photo2} conjectured that the radical driving this hydrogen abstraction chain is a CH$^\bullet_3$ generated by an earlier photolytic dissociation process.
This observation pinpoints the importance of accurately computing the activation energy for a reaction of the form $\mathrm{XCH_3 + CH_3^\bullet = XCH_2^\bullet + CH_4}$, where $\mathrm{XCH_3}$ denotes the DGEBA molecule, in order to understand the photodissociation process as a whole~\cite{mailhot2005study2}.

In this study, we consider a simplified structure possessing similar physical and chemical properties to DGEBA, specifically 
2,2-Diphenylpropane \cite{moleculeref}. While this simplified structure lacks the epoxide group present in DGEBA, we assume hydrogen abstraction from a central methyl group, the reaction of interest, is not significantly influenced by the presence or absence of the epoxide on the ends of the molecule.
To determine the reaction path of our radical model, we used density functional theory (DFT) with the B3LYP functional~\cite{Becke1993b3lyp} and D3 dispersion corrections~\cite{Grimme2010D3dispersion} at the  cc-PVDZ ~\cite{Dunning1989ccpvdz} level of theory as implemented in the Schr\"{o}dinger software~\cite{Schrodinger2024software}. 
We evaluated and optimized the energies of candidate reactant, transition state, and product geometries, obtaining the results in Fig.~\ref{fig:radical-epoxy-model}. 
We located the saddle point indicating the transition state using the Quadratic Synchronous Transit (QST) method as implemented in Schr\"{o}dinger. 
For the minima and transition-state searches, we fixed the distance between the carbon atoms of the CH$_3$ radical and the methyl group involved in hydrogen abstraction to $R_{\mathrm{C-C}} =2.69$ \AA.

\subsubsection{Classical pre-processing}

\begin{figure}[h!]
    \centering
\includegraphics[width=0.6\columnwidth]{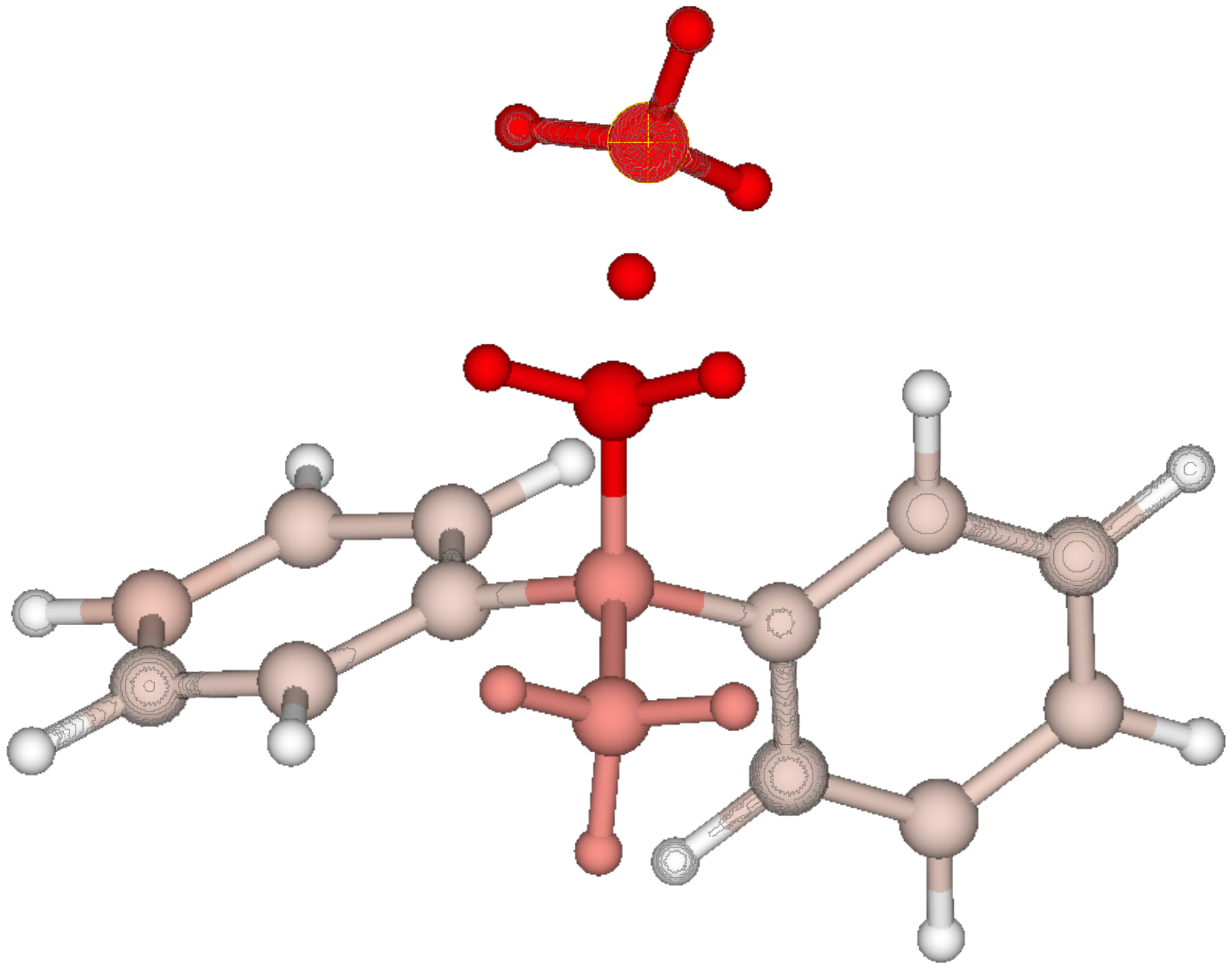}
    \caption{Active-space selection, exemplified for the transition-state geometry. Intrinsic bond orbitals (IBOs) are constructed, and those localized around the atoms in red (the $\mathrm{CH_3^\bullet}$ and $\mathrm{CH_4}$ where the hydrogen abstraction occurs) define the (13e,13o) active space. Similarly, IBOs localized around the atoms in red and dark pink define the (23e,23o) active space. The (39e,39o) active space is defined by expanding the (23e,23o) active space to include the pi-orbitals of the phenyl rings along with the sigma orbitals connecting the rings to the rest of the molecule.} \label{fig:ibo}
\end{figure}

We compute the activation energy $\Delta E^{\ddagger} = E_{\mathrm{TS}} - E_{\mathrm{R}}$ and the reaction energy $\Delta E = E_{\mathrm{P}} - E_{\mathrm{R}}$ in three different active spaces of sizes (13e,13o),  (23e,23o) and (39e,39o).
These active spaces are spanned by intrinsic bond orbitals~\cite{knizia2013intrinsic} located around $\mathrm{CH_3^\bullet}$ and $\mathrm{CH_4}$ where hydrogen abstraction occurs for the active space (13e,13o) and around additional neighboring atoms, as shown in Fig.~\ref{fig:ibo} for the active spaces (23e,23o) and (39e,39o). 
Both active spaces have one additional alpha electron compared to the beta electrons. The reason for considering three 
different active spaces is to assess the performance of the combination between SQD and EF across multiple system sizes.
In each active space, we used the classical methods restricted open-shell Hartree-Fock (ROHF), coupled-cluster singles and doubles (CCSD), and CCSD with perturbative triples (CCSD(T)), as implemented in the PySCF software package~\cite{sun2018pyscf,sun2020recent}.
We also used the heat-bath configuration interaction (HCI) as implemented in the DICE package~\cite{sharma2017semistochastic} -- working in the basis of CCSD natural orbitals and using truncation thresholds of $10^{-7}$, $5 \cdot 10^{-6}$, and $5 \cdot 10^{-5}$ a.u. for the active spaces -- and the density matrix renormalization group (DMRG) as implemented in the Block2 package~\cite{zhai2023block2} -- using a bond dimension of $4000$, SU(2) symmetry, and working in the basis of CCSD natural orbitals ordered by a genetic algorithm.

\subsubsection{Details of quantum simulations}

 
We constructed the EF and LUCJ quantum circuits using the open-source ffsim library~\cite{ffsim} and an in-house code to supply the circuit parameters based on classical CCSD and ResHF calculations.
We executed these quantum circuits on IBM's 133-qubit and 156-qubit Heron superconducting quantum processors $\mathsf{ibm\_torino}$ and $\mathsf{ibm\_kingston}$, respectively, gathering 500000 measurement outcomes for each circuit. 

To mitigate quantum errors, we use dynamical decoupling~\cite{viola1998dynamical,kofman2001universal,biercuk2009optimized} as available through the \texttt{SamplerV2} primitive of Qiskit's \texttt{Runtime} library where appropriate. 

After executing the EF quantum circuits, we collated the measurement outcomes from the individual circuits as discussed in Section~\ref{sec:combination}. 
For the (39e,39o) active space, we sub-sampled raw bitstrings from individual independent and superposition circuits that had a hamming weight difference of less than 10 from the Hartree-Fock bitstring for creating the collated dictionaries.  

We then used SQD to approximate molecular eigenstates, as implemented in IBM's SQD addon packages using DICE and GPU accelerated SBD solvers~\cite{sqd_addon, shirakawa2025closedloop, doi2026gpuacceleratedselectedbasisdiagonalization}. 

\vfill

\pagebreak
\newpage

\vfill

\pagebreak
\newpage

\section{Results}

\begin{figure*}[t]
    \centering
    \includegraphics[width=\columnwidth]{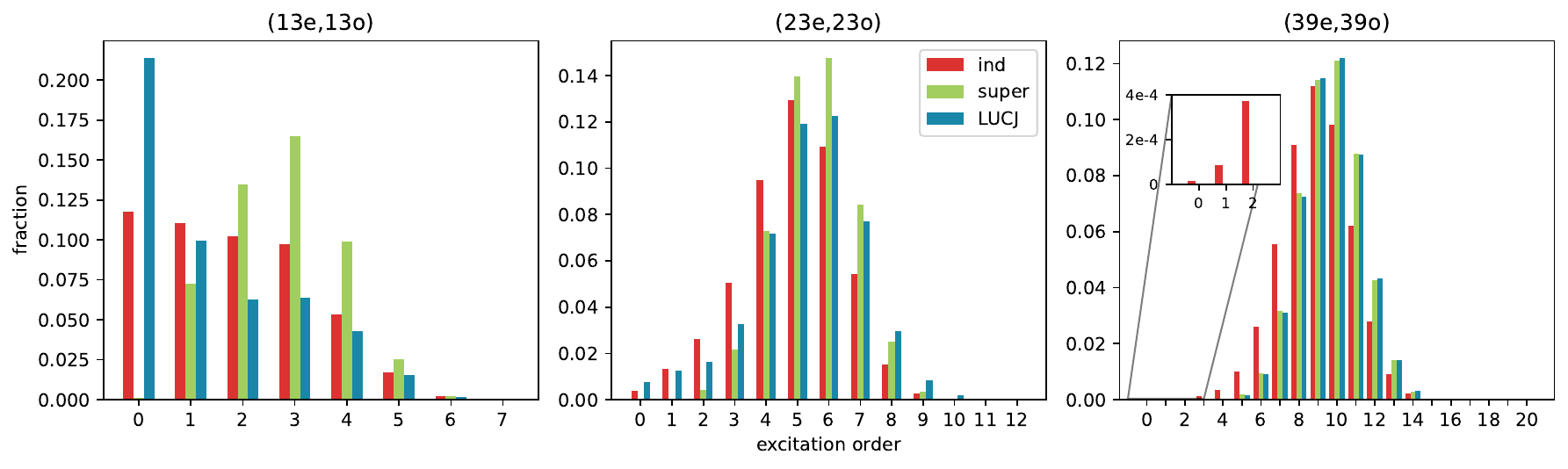}
    \caption{Fraction of measured bitstrings with the correct number of alpha and beta electrons on the quantum computer that contain the Hartree-Fock bitstring (excitation 0) and subsequent excitations (indexed by integers as the excitation order) on the Hartree-Fock bitstring for the three active space sizes for the individual (ind) circuits, superposition (super) and LUCJ circuits. The values are averaged over the geometries and electron spin.} \label{fig:excitations}
\end{figure*}
In Table~\ref{table:HWresources}, we provide information about the total number of qubits including ancilla qubits, two-qubit gates, and circuit depths for the EF and LUCJ circuits after transpilation to the HW. We see that the required 2 qubit gates and circuit depths are significantly smaller for the EF circuits when compared to the LUCJ circuits. 
This leads to higher noise in bitstrings obtained from the LUCJ circuits, especially as active spaces get larger. While the fraction of correct number of alpha and beta bitstrings is indicative of the relative quality of the sampled bitstrings on the hardware (HW), there are additional factors that determine that quality of the computed energies via the SQD procedure. Specifically, as seen in Fig.~\ref{fig:excitations}, we observe that the Hartree-Fock bitstring and low-lying excitations of the Hartree-Fock state are not present in the HW sampled bitstrings obtained from the LUCJ ansatz for the (39e,39o) active space due to impact of the HW noise even after applying dynamical decoupling. Further comparison of the energies obtained from the EF and the LUCJ ansatz are reported in Appendix \ref{sec:lucj_vs_EF}.

\begin{table}
\begin{tabular}{cccc}
\hline\hline
circuit & qubits  & 2-qubit gates & depth \\
  & (system + ancilla) & (cz) &  \\
\hline
(13e,13o) EF (ind) &13	&255 & 249\\
(13e,13o) EF (super) &13+1	& 450 & 475\\
(13e,13o) LUCJ &26+4 &877 &379 \\
\hline
(23e,23o) EF (ind) &23 & 800 & 435 \\
(23e,23o) EF (super) &23+1 &1375	& 834\\
(23e,23o) LUCJ &46+6 &2713 &655 \\
\hline
(39e,39o) EF (ind) &39 &2295 & 737\\
(39e,39o) EF (super) &39+1 &3894 & 1431\\
(39e,39o) LUCJ &78+5 &7655 & 1075\\
\hline\hline
\end{tabular}
\caption{Quantum resources to execute individual (ind) and superposition (super) EF circuits and standard LUCJ circuits after transpilation on devices with heavy-hex topology. Numbers reported for the two-qubit gates and depth are averaged over the spin-up and spin-down circuits (for EF) as well as geometries (for EF and LUCJ).}
\label{table:HWresources}
\end{table}


\begin{figure*}
    \centering
    \begin{subfigure}[b]{\columnwidth}
        \centering
        \includegraphics[width=\columnwidth]{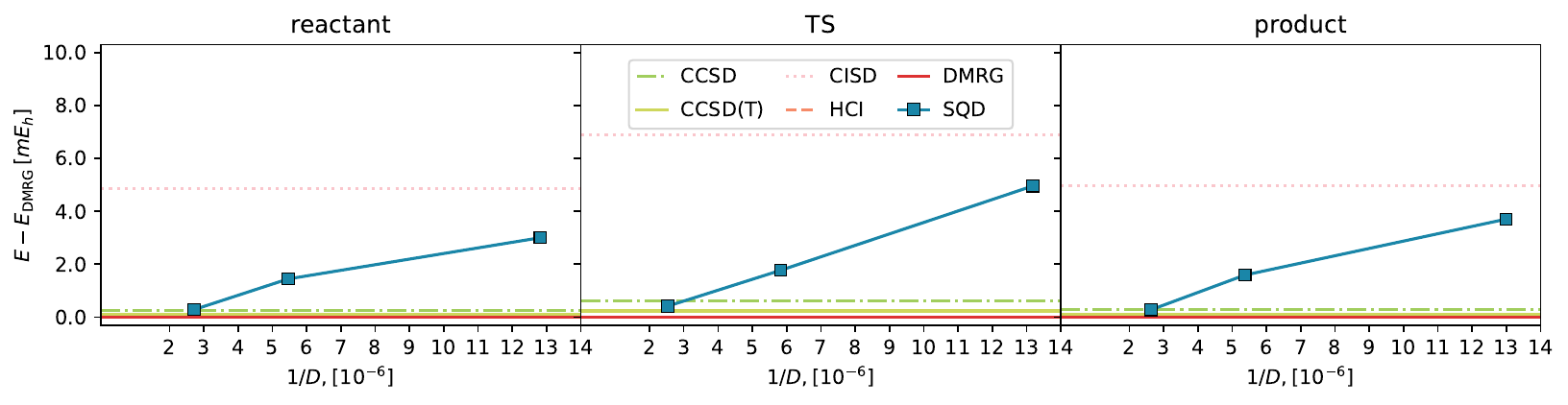}
        \caption{Deviation between SQD and DMRG energy (blue squares) as a function of the subspace dimension for reactant, transition state, and product (left to right), along with comparisons to other classical methods (horizontal lines).}
        \label{fig:1313}
    \end{subfigure}
    \hfill 
    \begin{subfigure}[b]{\columnwidth}
        \centering
        \includegraphics[width=\columnwidth]{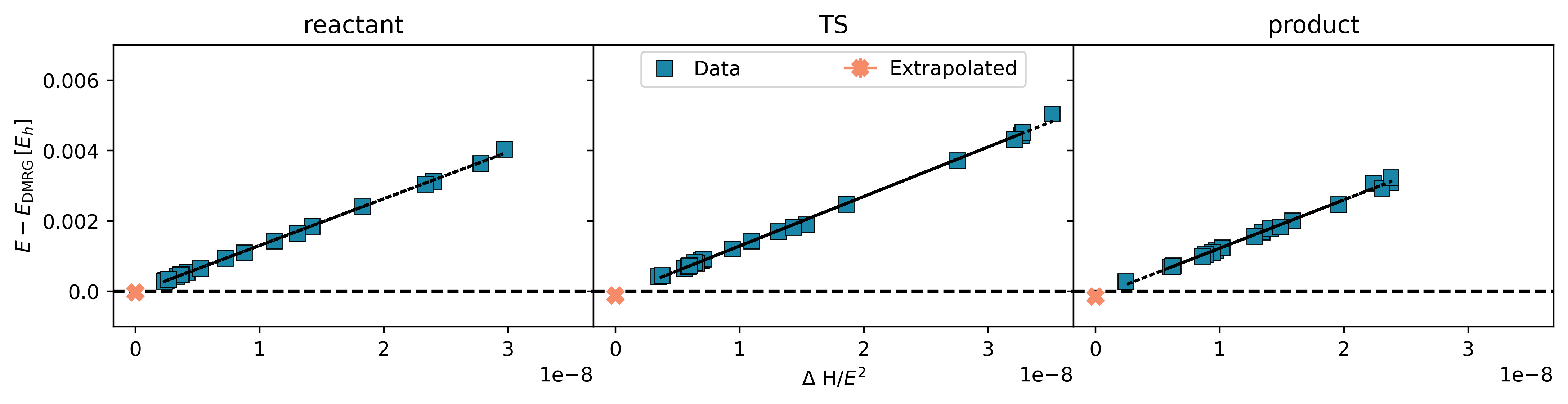}
        \caption{Energy vs variance linear extrapolation based on diagonalizations conducted during configuration recovery iterations for different batches with $10^{3}$ samples per batch.}
        \label{fig:1313_EV}
    \end{subfigure}
    
    \caption{SQD energies obtained in (13e,13o) active space using the Entanglement Forging ansatz.}
    \label{fig:1313_combined}
\end{figure*}

\begin{figure*}
    \centering
    \begin{subfigure}[b]{\columnwidth}
        \centering
        \includegraphics[width=\columnwidth]{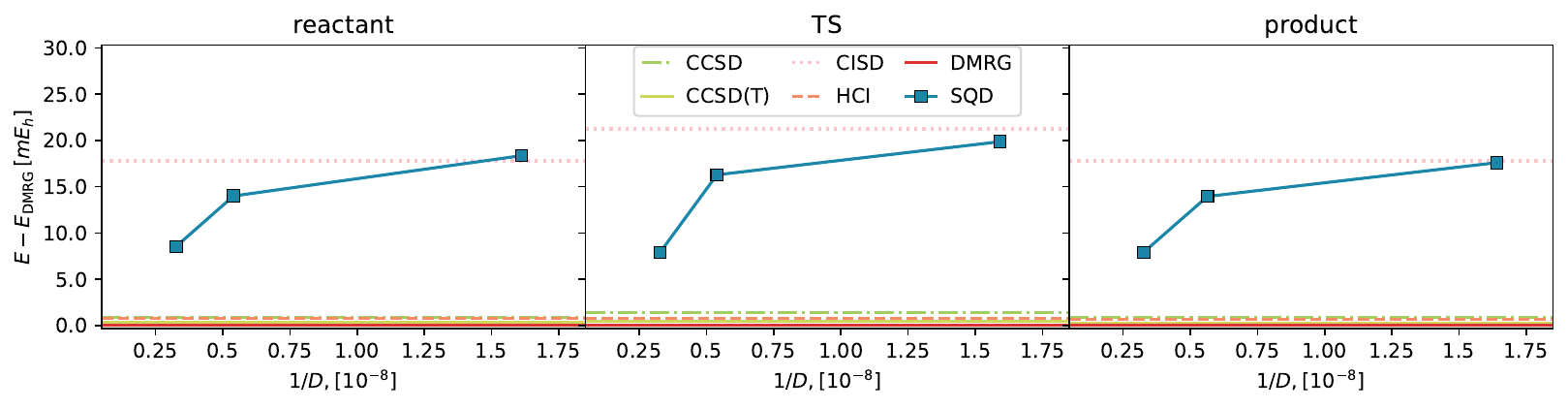}
        \caption{Deviation between SQD and DMRG energy (blue squares) as a function of the subspace dimension for reactant, transition state, and product (left to right), along with comparisons to other classical methods (horizontal lines).}
        \label{fig:2323}
    \end{subfigure}
    \hfill 
    \begin{subfigure}[b]{\columnwidth}
        \centering
        \includegraphics[width=\columnwidth]{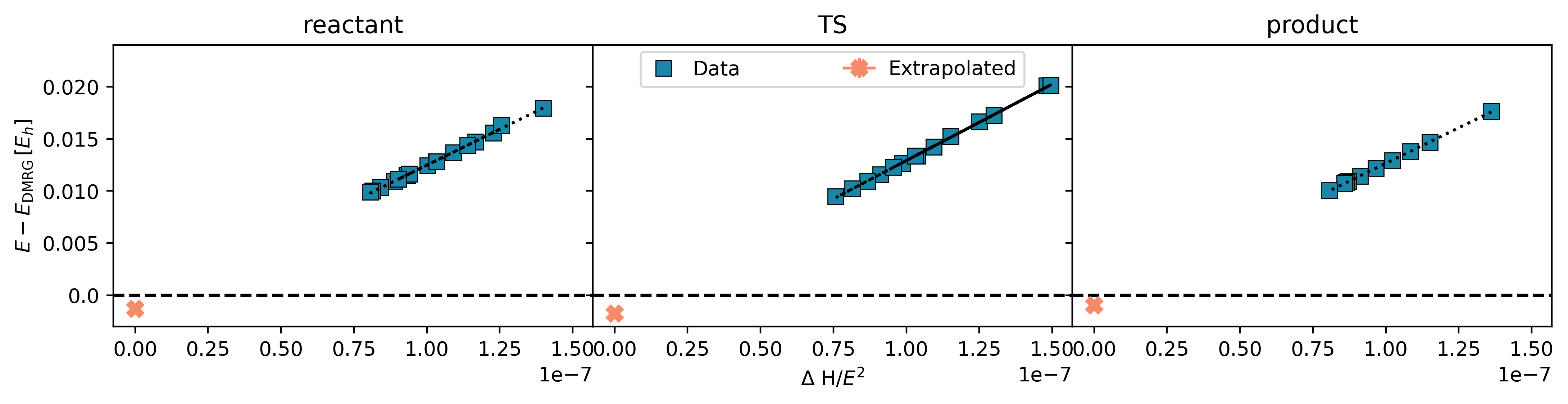}
        \caption{Energy vs variance linear extrapolation based on diagonalizations conducted during configuration recovery iterations for different batches with $4*10^{4}$ samples per batch.}
        \label{fig:2323_EV}
    \end{subfigure}
    
    \caption{SQD energies obtained in (23e,23o) active space using the Entanglement Forging ansatz.}
    \label{fig:2323_combined}
\end{figure*}

\begin{figure*}
    \centering
    \begin{subfigure}[b]{\columnwidth}
        \centering
        \includegraphics[width=\columnwidth]{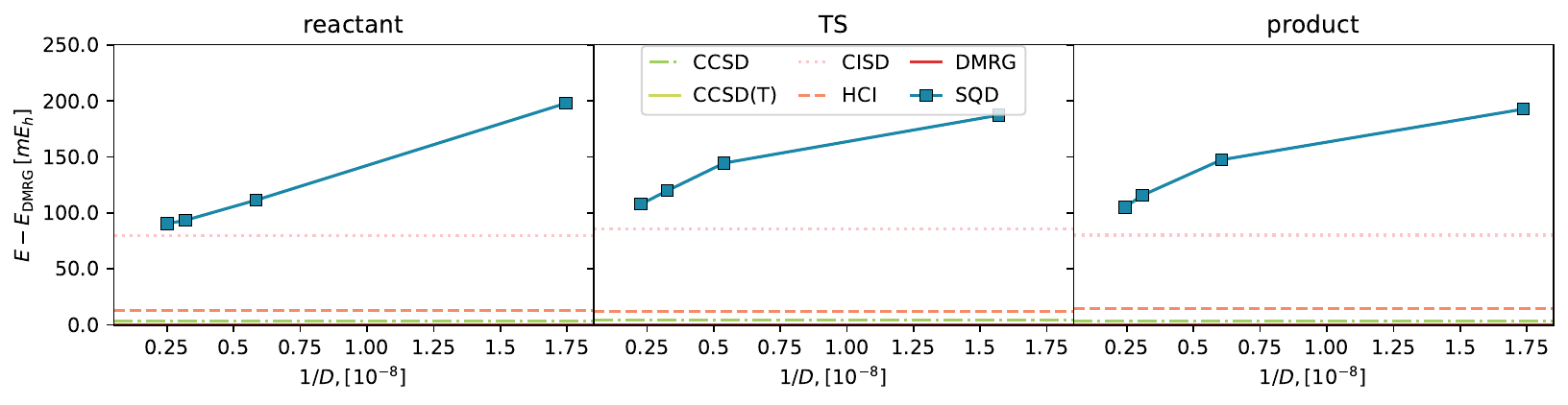}
        \caption{Deviation between SQD and DMRG energy (blue squares) as a function of the subspace dimension for reactant, transition state, and product (left to right), along with comparisons to other classical methods (horizontal lines).}
        \label{fig:3939}
    \end{subfigure}
    \hfill 
    \begin{subfigure}[b]{\columnwidth}
        \centering
        \includegraphics[width=\columnwidth]{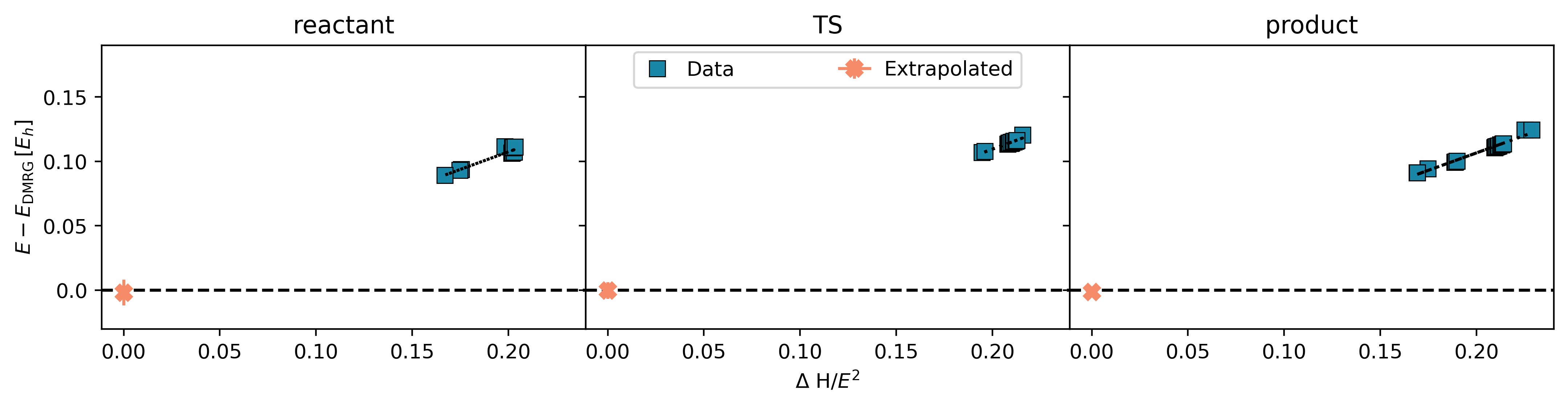}
        \caption{Energy vs variance linear extrapolation based on diagonalizations conducted during configuration recovery iterations for different batches.}
        \label{fig:3939_EV}
    \end{subfigure}
    
    \caption{SQD energies obtained in (39e,39o) active space using the Entanglement Forging ansatz.}
    \label{fig:combined}
\end{figure*}



For each geometry we run $K$ independent SQD calculations also called batches with multiple $N_{spb}$ samples per batch and $n_{CR}$ self-consistent configuration recovery iterations. We set the carryover threshold to be $10^{-4}$ in all the calculations. Minimum energy across iterations and batches are reported for different values of $N_{spb}$.

In Fig.~\ref{fig:1313}, we show the ground-state energy of reactant, transition state, and product in the (13e,13o) active space. We used $K=5$, $n_{CR}=5$ and $N_{spb} = \{250,500,1000\}$. The SQD energy is computed by diagonalizing the Hamiltonian in subspaces spanned by $D$ configurations which is indirectly controlled by $N_{spb}$. As expected, the SQD energy decreases as $N_{spb}$ increases, since increasing $N_{spb}$ leads to higher $D$. For all values of $N_{spb}$ we studied, the SQD energy is below CISD and, for $N_{spb} \simeq 1000$ which gives a subspace dimension of $D \simeq 3.8*10^{5}$, it agrees well with CCSD, CCSD(T), HCI and DMRG within 1m$E_{h}$. The predicted energies at zero variance obtained via linear extrapolation of energies vs variances, shown in Fig.~\ref{fig:1313_EV}, is also within 1m$E_{h}$ of the DMRG values for each geometry. For comparison, we also show the energy at zero variance obtained from linear extrapolation of energy vs variance calculated as $\frac{\Delta H}{E^{2}} = \frac{\langle H^{2}\rangle - \langle H \rangle^{2}}{E^{2}}$ in Fig.~\ref{fig:1313_EV}.

In Fig.~\ref{fig:2323}, we now consider results for the (23e,23o) active space. Here, we use $N_{spb} = \{10^{4},2*10^{4},4*10^{4}\}$. For the smaller two values of $N_{spb}$, we use $K=5$ and $n_{CR}=5$; whereas, $K=3$ with $n_{CR}=10$ is used for the largest value. As seen in this figure, the minimum SQD energy obtained is less than 10m$E_{h}$ above the reference DMRG value with a subspace dimension of $D \simeq 3.06*10^{8}$. The energy obtained at zero variance from the energy-variance linear extrapolation in Fig.~\ref{fig:2323_EV} is again within 1.8m$E_{h}$ of the DMRG values for each of the geometries.

For the largest active space studied, (39e,39o), we report the results obtained by performing SQD computations with $N_{spb} = \{2.5*10^{4},4.5*10^{4},6*10^{4},10^{5}\}$. For the smaller subspace sizes we used $K=5$ and $n_{CR}=5$. For the largest values of $N_{spb}$, we used $K={1}$ and $n_{CR}=10$. In addition to carryover method being used within a run with given $N_{spb}$, we also applied carryover of bitstrings and occupation numbers across different $N_{spb}$ computations. Specifically, important bitstrings from lowest energy SQD run with a smaller $N_{spb}$ calculation was carried over to the subsequent larger $N_{spb}$ runs and occupations numbers from the best SQD run was used as input for the larger runs. Fig.~\ref{fig:3939}, shows that the lowest energies obtained from the SQD runs is roughly around the CISD values and about 100m$E_{h}$ above the DMRG reference values. These results were obtained with a subspace dimension of $D \simeq 4.3*10^{8}$. While there is a large energy error compared to the DMRG values, this gap closes when we look at the energies obtained at zero variance from the linear energy-variance extrapolation plots. In Fig.~\ref{fig:3939_EV}, we see that the extrapolated values are less than 2m$E_{h}$ away from the DMRG benchmark across the three system. This result can be improved further with more runs for larger samples per batch calculations.

Energies across all the systems for selected classical methods, SQD and extrapolated SQD values are reported in \ref{table:energy_data} in the Appendix.

In Fig.~\ref{fig:de} we consider the activation energy $E_{\mathrm{TS}}-E_{\mathrm{R}}$ and the reaction energy $E_{\mathrm{P}}-E_{\mathrm{R}}$. As seen, CCSD, CCSD(T), HCI, and DMRG agree well within 2 kcal/mol for these energy differences in all the active spaces considered. It is important to note that HCI predicts that the product is lower in energy than the reactant for the (39e,39o) active space. In comparison, RHF, and CISD overestimate the activation energy for all active spaces.
The performance of extrapolated SQD method visibly dependent on the active-space size: 
while the mean values of the extrapolated energies are within 1kcal/mol of the DMRG values for all the active space sizes, the error bars associated with these extrapolations become larger as active space sizes increase. These error bars can be controlled further by increasing the subspace dimensions of the diagonalization space as well as gathering more statistics. 


\begin{figure}[h!]
    \centering
\includegraphics[width=\columnwidth]{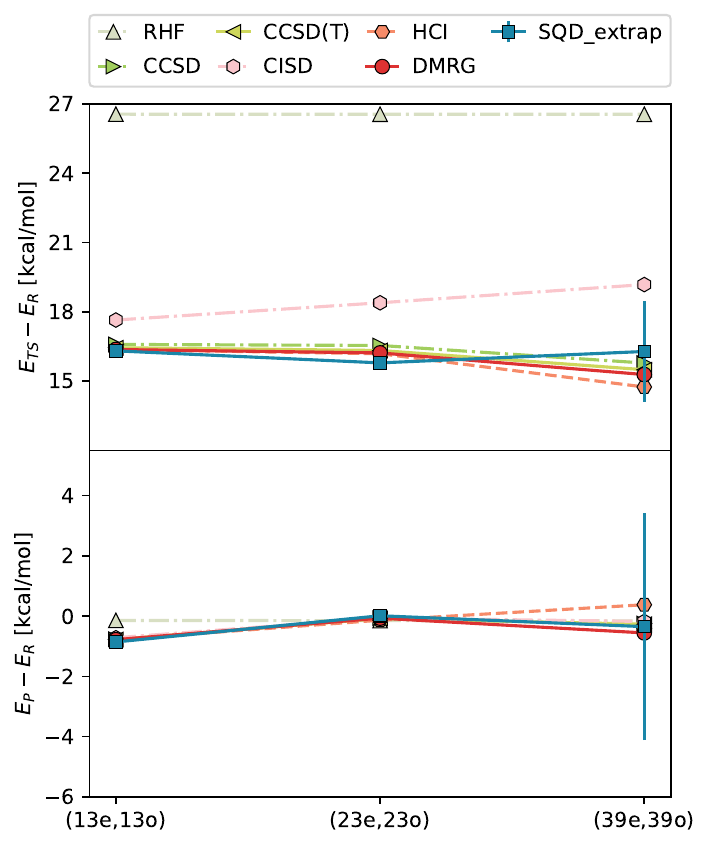}
    \caption{Activation energy $E_{\mathrm{TS}}-E_{\mathrm{R}}$ and reaction energy $E_{\mathrm{P}}-E_{\mathrm{R}}$ from classical methods and extrapolated SQD, using active spaces of (13e,13o), (23e,23o) and (39e,39o).}
    \label{fig:de}
\end{figure}

\section{Conclusion and Outlook}

In this work, we combined the entanglement forging (EF) and sample-based quantum diagonalization (SQD) methods.
The main difficulty posed by this integration is that the probability distribution of the electronic configuration over an EF wavefunction, Eq.~\eqref{eq:ef_pdf}, is not a compound probability distribution. To overcome this difficulty, we proposed to sample configurations from an approximate compound probability distribution, Eq.~\eqref{eq:ef_pdf_compound}.
To increase the flexibility of EF, both in variational and SQD simulations, we proposed a more general family of EF wavefunction, wherein linear combinations of nonorthogonal Slater determinants and an approximate unitary coupled-cluster wavefunction (called local unitary cluster Jastrow or LUCJ) replace the ``bitstrings'' (i.e., individual electronic configurations) and hopgates (i.e., two-qubit gates) used in previous studies. The choice of non-orthogonal Slater determinants and an approximate unitary coupled-cluster wavefunction is motivated by a connection between EF and coupled-cluster theory, revealed by a Hubbard-Stratonovich transformation.

The quantum circuits required by the combination of SQD and EF are divided into: diagonal (for the preparation of a Slater determinant and the application of an LUCJ circuit) and superposition (for the preparation of a linear combination of two Slater determinants and the application of an LUCJ circuit). The former have $\sim 50\%$ the qubits and quantum gates of an equivalent LUCJ circuit for alpha-beta spins considered together and slightly lower depth. The latter, due to an economization of the ``Hadamard test'' circuit for the preparation of superpositions of Slater determinants, have $(M+1)(M-1)$ additional gates and $2(M-1)$ more layers of two-qubit gates. As such, they are always deeper than a conventional LUCJ circuit and have fewer gates for $L \geq 2$ repetitions. We note that EF circuits -- having half the qubits than conventional circuits and requiring linear connectivity in this case -- may be executed in parallel and benefit from more freedom in the choice of qubit layout.

We used the combination of EF and SQD to simulate the activation energy and reaction energy of a hydrogen abstraction reaction motivated by photodegradation of composite polymers, in active spaces of (13e, 13o), (23e, 23o) and (39e,39o) to assess the performance of the method with increasing numbers of active electrons and orbitals.
As the electronic wavefunctions along the steps of the reaction are dominated by the mean-field configuration (i.e. electron correlation is purely dynamical), we used different classical methods -- CCSD, CCSD(T), HCI, and DMRG -- to provide reference results against which to compare EF.
For the (13e,13o), (23e,23o) and (39e,39o) active space, SQD + EF provided results of accuracy comparable to reference classical methods, see Fig.~\ref{fig:1313_combined} and Fig.~\ref{fig:2323_combined}. In comparison, LUCJ fails to produce meaningful diagonalization subspaces for the (39e,39o) active space due to excess noise in the sampled bitstrings form the hardware. 

This work may be continued and refined in different ways. First, by optimizing free parameters in the EF wavefunction -- both in quantum circuits and in the probabilities $P_I$ in Eq.~\eqref{eq:ef_pdf_compound} -- to minimize the SQD energy~\cite{kanno2023quantum}. Second, by moving beyond hydrogen abstraction towards other molecular species, including with multireference electronic character, e.g., to perform a systematic assessment of accuracy. Finally, the techniques presented here may be a starting point to combine SQD with circuit knitting techniques~\cite{piveteau2023circuit}.

\section*{Acknowledgments}
The authors thank Gavin Jones, Robert Walkup, Sophia Wen and Seetharami Seelam for insightful discussions and support with SBD software.

\appendix

\section{Additional details about EF}
\label{sec:appendix_gef}

In this Section, we derive Eq.~\eqref{eq:sum_of_squares} and Eq.~\eqref{eq:uCCSD_forging} of the main text, and provide more information about the parameterization of Eq.~\eqref{eq:gef_state}.

Let us start by performing a singular-value decomposition (SVD) of the opposite-spin $t_2$ amplitudes in Eq.~\eqref{eq:t_op},
\begin{equation}
\begin{split}
\op{T}_{\alpha\beta} 
&= \sum_{aiBJ} (t_{2 \alpha \beta})_{aiBJ} \op{E}^\alpha_{ai} \op{E}^\beta_{BJ} \\
&= \sum_s \tau_s \, \sum_{aiBJ} U^s_{ai} V^s_{BJ} \op{E}^\alpha_{ai} \op{E}^\beta_{BJ} \\
&= \sum_s \left[ \sum_{ai} \sqrt{ \tau_s } \, U^s_{ai} \, \op{E}^\alpha_{ai} \right] \left[ \sqrt{ \tau_s } \, \sum_{BJ} V^s_{BJ} \, \op{E}^\beta_{BJ} \right] \\
&= \sum_s \op{u}_s \op{V}_s = \sum_s \frac{( \op{u}_s + \op{V}_s )^2 + ( i \op{u}_s - i \op{V}_s )^2}{4} \;.
\end{split}
\end{equation}
Since the $\op{u}_s$ and $\op{V}_s$ commute, we can write
\begin{equation}
\begin{split}
\op{T}_{\alpha\beta} - \op{T}_{\alpha\beta}^\dagger
&= \frac{1}{4} \sum_s ( \op{u}_s + \op{V}_s )^2 + ( i \op{u}_s - i \op{V}_s )^2 \\
&+ \frac{1}{4} \sum_s ( i \op{u}^\dagger_s + i \op{V}^\dagger_s )^2 + ( \op{u}^\dagger_s - \op{V}^\dagger_s )^2
\end{split}
\end{equation}
and this equation defines the squares of one-body operators in Eq.~\eqref{eq:sum_of_squares}. To prove Eq.~\eqref{eq:uCCSD_forging}, 
we consider the following Trotter approximation
\begin{equation}
e^{\op{T} - \op{T}^\dagger} \simeq 
e^{\op{T}_{\alpha\alpha} - \op{T}_{\alpha\alpha}^\dagger}
e^{\op{T}_{\beta\beta} - \op{T}_{\beta\beta}^\dagger}
\prod_\delta e^{\frac{\op{X}_\delta^2}{2}}
\;.
\end{equation}
We then use the Hubbard-Stratonovich representation to obtain
\begin{equation}
e^{\op{T} - \op{T}^\dagger} \simeq 
e^{\op{T}_{\alpha\alpha} - \op{T}_{\alpha\alpha}^\dagger}
e^{\op{T}_{\beta\beta} - \op{T}_{\beta\beta}^\dagger}
\prod_\delta \int_{-\infty}^{\infty} dy_\delta \, \frac{e^{- \frac{y_\delta^2}{2}}}{\sqrt{2\pi}} \, e^{y_\delta \op{X}_\delta}
\;.
\end{equation}
Let us now introduce the approximation
\begin{equation}
\prod_\delta e^{y_\delta \op{X}_\delta}
\simeq 
e^{ \sum_\delta y_\delta \op{X}_\delta}
\end{equation}
and the definitions $\bts{y} = \{ y_\delta \}_\delta$ and 
$p(\bts{y}) = \prod_\delta \frac{e^{- \frac{y_\delta^2}{2}}}{\sqrt{2\pi}}$,
obtaining
\begin{equation}
e^{\op{T} - \op{T}^\dagger} \simeq 
e^{\op{T}_{\alpha\alpha} - \op{T}_{\alpha\alpha}^\dagger}
e^{\op{T}_{\beta\beta} - \op{T}_{\beta\beta}^\dagger}
\prod_\delta \int d\bts{y} \, p(\bts{y}) \, e^{ \sum_\delta y_\delta \op{X}_\delta}
\;,
\end{equation}
from which Eq.~\eqref{eq:uCCSD_forging} immediately follows.

To define the wavefunctions in Eq.~\eqref{eq:gef_state}, we use a two-step procedure.
\begin{itemize}
\item First, we perform a CCSD calculation yielding amplitudes $t_{2\alpha\alpha}$, $t_{2\beta\beta}$, and $t_{2\alpha\beta}$, and we use the same-spin amplitudes to parametrize the LUCJ circuit as detailed in the supplementary information of Ref.~\cite{robledo2024chemistry}.
\item Second, we perform a ResHF calculation to define the determinants $d_{\mu\sigma}$ and coefficients $c_\mu$. We solve the ResHF equations as defined in Ref.~\cite{fukutome1988theory}, using the following ``warm-starting'' procedure to provide initial orbitals and coefficients:
\begin{enumerate}
\item we choose a number of determinants $N_{\mathrm{det}}$, 
\item we sample $N_{\mathrm{det}}$ auxiliary fields $\{ \bts{y}_\mu \}_{\mu=1}^{N_{\mathrm{det}}}$ from the normal distribution in Eq.~\eqref{eq:uCCSD_forging} defining Slater determinants $| \Phi(\bts{y}_\mu) \rangle$, 
\item we variationally optimize $\sum_\mu c_\mu | \Phi(\bts{y}_\mu) \rangle$ with respect to $c_\mu$ and $\bts{y}_\mu$. More specifically, we define $\tilde{c}_\mu(\bts{y})$ as the lowest-energy solution of $\sum_\mu \langle \Phi(\bts{y}_\nu) | \hat{H} | \Phi(\bts{y}_\mu) \rangle \, \tilde{c}_\mu(\bts{y}) = E_0(\bts{y})$ $\sum_\mu \langle \Phi(\bts{y}_\nu) | \Phi(\bts{y}_\mu) \rangle \, \tilde{c}_\mu(\bts{y})$ and minimize $E_0(\bts{y})$ as a function of $\bts{y}$. We use the resulting linear combination of non-orthogonal determinants as the initial point of the variational optimization in  Ref.~\cite{fukutome1988theory}.
\end{enumerate}
\end{itemize}

\section{Additional details about quantum circuits}

In this Section, we show that the circuits in Fig.~\ref{fig:superposition_circuits}$c$ prepare the superposition state in Eq.~\eqref{eq:superposition_of_dets}.
First, a register of qubits prepared in $| d_{\mu \alpha} \rangle$ is coupled to an ancilla prepared in the state $| \varphi_p \rangle = \frac{ | 0 \rangle + i^p | 1 \rangle }{ \sqrt{2} }$. Then, the controlled operation
\begin{equation}
\mathsf{c}[ e^{\op{X}^{\mu\nu}_\alpha} ] = | 0 \rangle \langle 0 | \otimes I + | 1 \rangle \langle 1 | \otimes e^{\op{X}^{\mu\nu}_\alpha}
\end{equation}
is applied, producing the output state
\begin{equation}
\begin{split}
| 0 \rangle | d_{\mu\alpha} \rangle + i^p | 1 \rangle | d_{\nu \alpha} \rangle &= | + \rangle \left[ \frac{ | d_{\mu\alpha} \rangle + i^p | d_{\nu \alpha} \rangle}{\sqrt{2}} \right] \\
&+ | - \rangle \left[ \frac{ | d_{\mu\alpha} \rangle - i^p | d_{\nu \alpha} \rangle}{\sqrt{2}} \right] \;.
\end{split}
\end{equation}
Recalling Eq.~\eqref{eq:superpositions}, we can write this equation as
\begin{equation}
| 0 \rangle | d_{\mu\alpha} \rangle + i^p | 1 \rangle | d_{\nu \alpha} \rangle 
=
\frac{U_{\mu\nu}^p}{\sqrt{2}} | + \rangle | d_{\mu\nu \alpha}^p \rangle
+ 
\frac{U_{\mu\nu}^{p+2}}{\sqrt{2}} | - \rangle | d_{\mu\nu \alpha}^{p+2} \rangle
\end{equation}
The ancilla is then measured in the Pauli $X$ basis, yielding results $\pm 1$ with probabilities
\begin{equation}
p(1) = \frac{1}{2} \left[ U_{\mu\nu}^p \right]^2 
\;,\;
p(-1) = \frac{1}{2} \left[ U_{\mu\nu}^{p+2} \right]^2 
\end{equation}
and collapsing the main register in the states $| d_{\mu\nu \alpha}^p \rangle$ and $| d_{\mu\nu \alpha}^{p+2} \rangle$ respectively. Note that, while the preparation of $| d_{\mu\nu \alpha}^p \rangle$ is probabilistic,
measurement outcomes need not be discarded, because the EF method also requires the preparation of $| d_{\mu\nu \alpha}^{p+2} \rangle$.

So far, we verified that the circuit in the top portion of Fig.~\ref{fig:superposition_circuits}$c$ prepares the state $| d_{\mu\nu \alpha}^p \rangle$. The controlled operation $\mathsf{c}[ e^{\op{X}^{\mu\nu}_\alpha} ]$,
if implemented naively, can be considerably expensive, leading to a circuit of $O(M^2)$ controlled two-qubit gates requiring all-to-all qubit connectivity. To achieve a more economic implementation, we recall that
$\op{X}^{\mu\nu}_\alpha = \sum_{pr} X^{\mu\nu}_{pr} \crt{p\alpha} \dst{r\alpha}$ is an anti-Hermitian one-body operator, and thus can be diagonalized by an orbital rotation $\hat{W} = \sum_{pr} W_{pr} \crt{p\alpha} \dst{r\alpha}$ 
with $\sum_r X^{\mu\nu}_{pr} W_{rq} = i \xi_q W_{pq}$. We can then write
\begin{equation}
\mathsf{c}[ e^{\op{X}^{\mu\nu}_\alpha} ] = \hat{W}^\dagger \mathsf{c}[ e^{\sum_q i \xi_q \crt{q\alpha} \dst{q\alpha}} ] \hat{W} \;.
\end{equation}
Since, in the Jordan-Wigner representation, $\crt{q\alpha} \dst{q\alpha} = |1 \rangle \langle 1|_q$, the controlled operation $\mathsf{c}[ e^{\sum_q i \xi_q \crt{q\alpha} \dst{q\alpha}} ]$ is a product of $M$ controlled single-qubit Z rotations.
While these controlled operations require all-to-all connectivity, we can easily implement them on a device with linear qubit connectivity by swapping the ancilla from the first to the last position of the register, i.e.
\begin{equation}
\mathsf{c}[ e^{\op{X}^{\mu\nu}_\alpha} ] = \hat{W}^\dagger \prod_{q=1}^M \mathsf{c}_{q-1} \mathsf{SWAP}_{q-1,q} \mathsf{c}_{q-1}[ e^{i \xi_q |1 \rangle \langle 1|}_q ] \hat{W} \;.
\end{equation}
as shown in Fig.~\ref{fig:superposition_circuits}$c-d$, where a circle traversed by a vertical line delimited by two crosses denotes a gate $\mathsf{SWAP}_{q-1,q} \mathsf{c}_{q-1}[ e^{i \xi_q |1 \rangle \langle 1|}_q ]$.

\section{Data of obtained EF energies}
A data table listing the energies obtained for selected methods for the reactant, TS and product across the different active space sizes are listed in table \ref{table:energy_data}. The reported values for HCI and SQD are the lowest obtained energies across different parameter runs.

\begin{table*}
\begin{tabular}{cccccc}
\hline\hline
systems (13e,13o)        & CCSD(T) & CASCI & DMRG & SQD  &SQD Extrap\\
\hline
reactant  & -617.0327472 & -617.0328557  &-617.0328557 & -617.0325708 & -617.0328845 +/- 0.0000128\\
TS        & -617.0065427 & -617.0067752  &-617.0067752 & -617.0063583 & -617.0068957 +/- 0.0000284\\
product   & -617.0339864 & -617.0341024  &-617.0341024 & -617.0338231 & -617.0342531 +/- 0.0000311\\
\hline
systems (23e,23o)        & CCSD(T) & HCI & DMRG & SQD  & SQD Extrap\\
\hline
reactant & -617.1301342 & -617.1296206 & -617.130429 & -617.1137078 & -617.1315923 +/- 0.0001385\\
TS       & -617.1041188 & -617.1038285 & -617.1045815& -617.0851395 & -617.1064329 +/- 0.0000884\\
product  & -617.1302647 & -617.1298376 & -617.1305384& -617.1129711 & -617.1315747 +/- 0.0000802\\
\hline
systems (39e,39o)        & CCSD(T) & HCI & DMRG & SQD  & SQD Extrap\\
\hline
reactant  & -617.3275864 & -617.3145993 & -617.3273225 & -617.2381694 & -617.3291567 +/- 0.0099048\\
TS        & -617.3028990 & -617.2911036 & -617.3029691 & -617.1960037 & -617.3032112 +/- 0.0064286\\
product   & -617.3280165 & -617.3139983 & -617.3282124 & -617.2236767 & -617.3297149 +/- 0.0038984\\
\hline\hline


\end{tabular}
\caption{Absolute energies for the reactant, TS and product system using different active space sizes of (13e,13o), (23e,23o) and (39e,39o) calculated using selected methods.}
\label{table:energy_data}
\end{table*}

\label{sec:lucj_vs_EF}

\section{Comparison of energies obtained from EF vs. LUCJ ansatz}
\label{sec:lucj_vs_EF}

\begin{figure*}
    \centering
    \includegraphics[width=\textwidth]{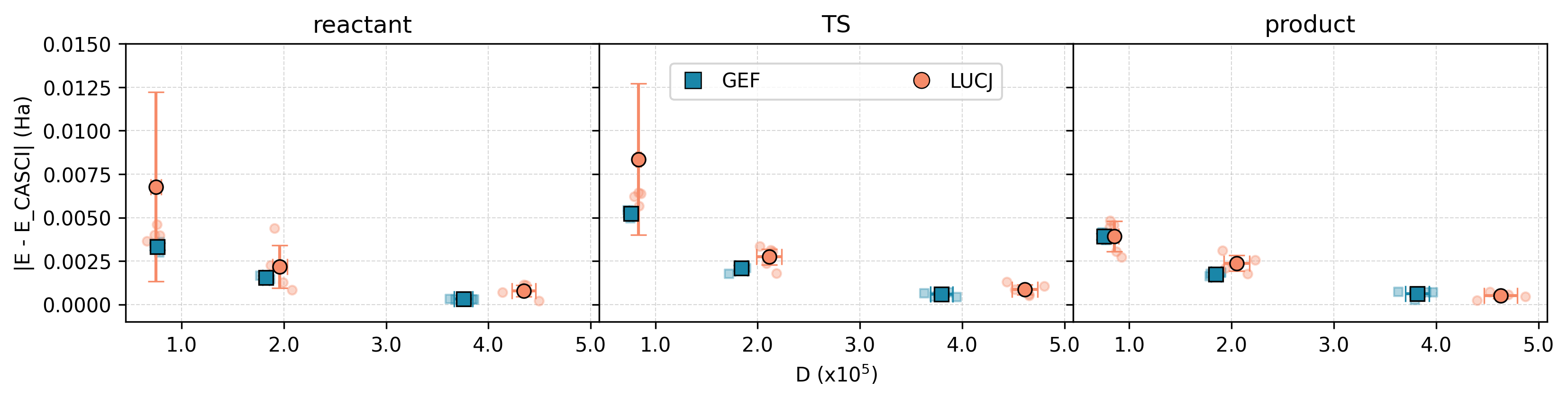}
    \caption{Deviation between SQD and CASCI energy as a function of the subspace dimension for reactant, transition state, and product (left to right) in a (13e,13o) active space. Squares and circles correspond to EF and LUCJ results, respectively, with error bars representing one standard deviation in both energy and subspace dimension.} \label{fig:1313_LUCJ_EF}
\end{figure*}

Entanglement Forging (EF) as described in this manuscript and the Local Unitary Cluster Jastrow Ansatz (LUCJ) are structurally distinct quantum circuit ansatze, both used within the Sample-based Quantum Diagonalization (SQD) framework to sample electronic configurations. 
Their relative performance is assessed through convergence of the computed electronic energy with subspace dimension (D). Energies are plotted as a function of D for the reactant (R), transition state (TS), and product (P) geometries. In Figure \ref{fig:1313_LUCJ_EF}, scatter points show energies of independent batch runs clustered as per the values of $N_{spb}$. Filled squares and circles denote EF and LUCJ means, respectively, with error bars representing one standard deviation in both energy and D. For both methods, increasing D monotonically lowers the energy, consistent with systematic improvement of the variational approximation.

For the (13e, 13o) reactant, EF consistently reaches energies closer to the CASCI reference at comparable D values, while LUCJ shows larger fluctuations and slower convergence, particularly at smaller shot counts and subspace dimensions. A similar trend is observed for the transition state, where EF achieves lower errors across all shot counts and LUCJ exhibits notably larger statistical dispersion, especially at 250 samples per batch (which correspond to lowest subspace dimensions studied here). For the product geometry, EF and LUCJ perform comparably in terms of accuracy. Overall, EF achieves equivalent or better accuracy for smaller sub-space dimensions compared to the LUCJ ansatz. 

\begin{figure*}
    \centering
    \includegraphics[width=\textwidth]{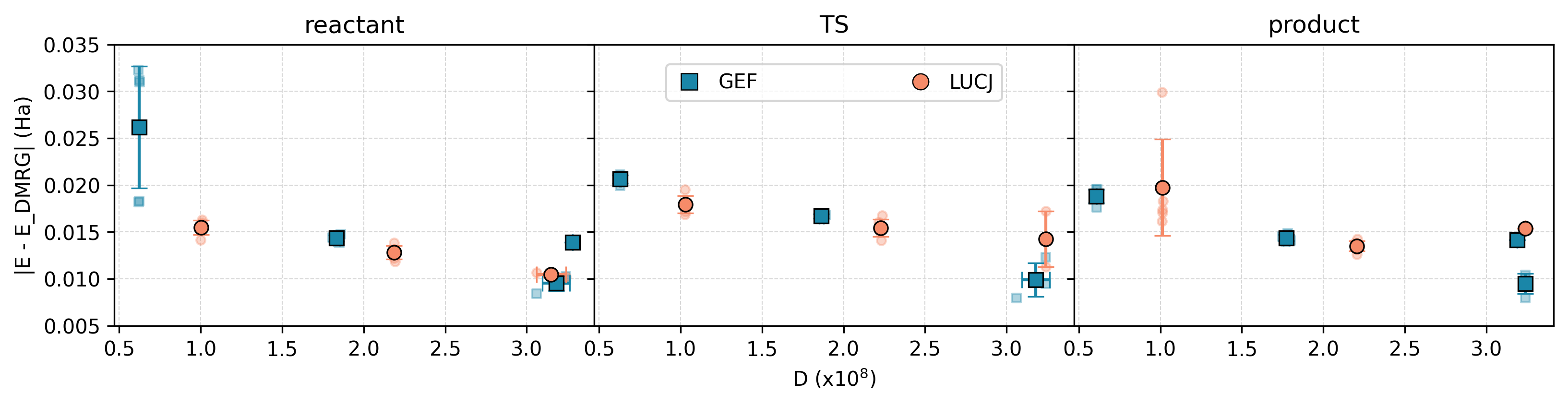}
    \caption{Deviation between SQD and DMRG energy as a function of the subspace dimension for reactant, transition state, and product (left to right) in a (23e,23o) active space. Squares and circles correspond to EF and LUCJ results, respectively, with error bars representing one standard deviation in both energy and subspace dimension.} \label{fig:2323_LUCJ_EF}
\end{figure*}

For the (23e, 23o) case (Figure \ref{fig:2323_LUCJ_EF}), energies are compared against the DMRG reference across the reactant, transition state, and product geometries, using samples per batch of 10k, 20k, 30k, and 40k for EF and 10k, 15k, and 20k for LUCJ. Both methods show systematic improvement with increasing subspace dimensions. We find that for equivalent subspace dimensions of about $3*10^{8}$, EF energies are lower than the LUCJ energies on average. 


We were not able to get equivalent results for the (39e,39o) active space since the raw HW counts for the LUCJ ansatz did not present the Hartree-Fock bitstring or any low lying excitations on the Hartree-Fock state. Configuration recovery was not effective in this case. Entanglement Forging thus allows us to access active spaces which would otherwise be inaccessible via the SQD method using the LUCJ ansatz.

\section{Additional details about the target reaction}
\label{sec:dgeba}

\begin{figure*}
     \centering
     \includegraphics[width=0.75\textwidth]{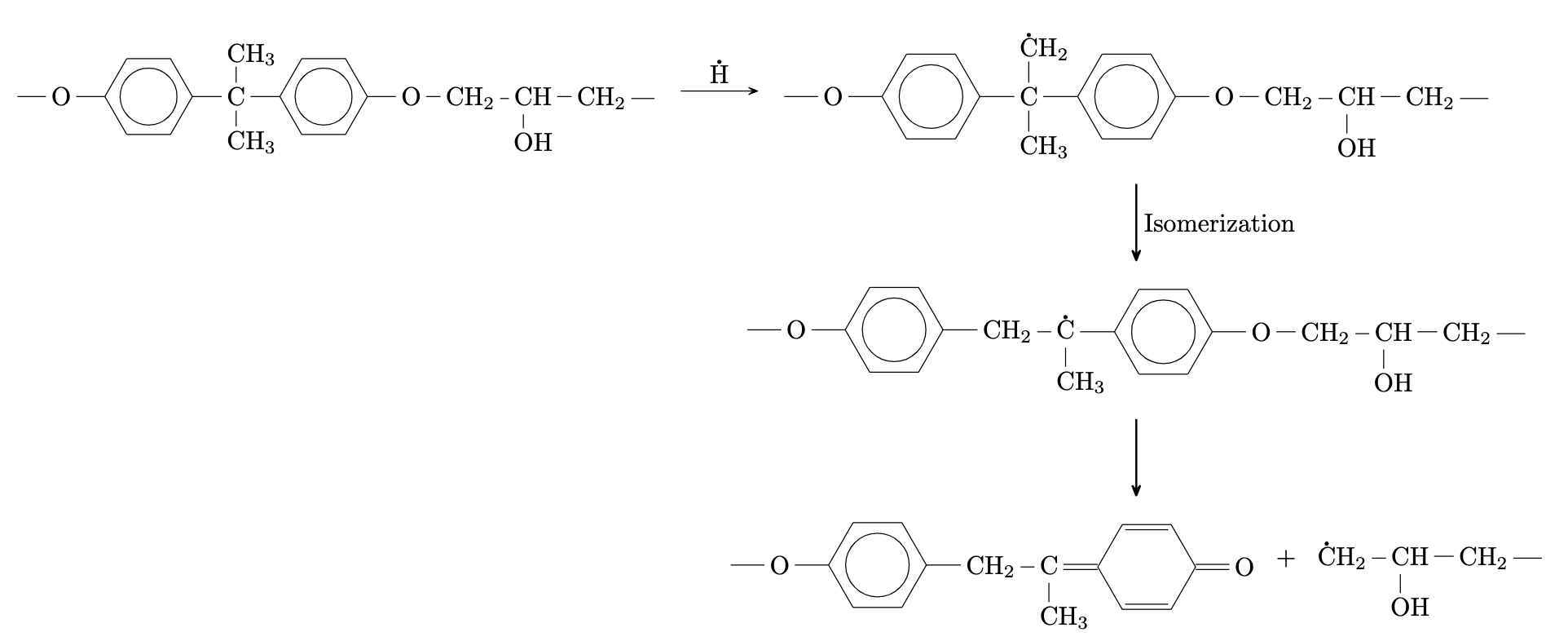}
     \caption{Key photodegradation reactions for DGEBA epoxy resin.}
     \label{fig:all-reaction-steps}
\end{figure*} 

This study is motivated by the photo-oxidation of the epoxy resin DGEBA~\cite{raanby1989photodegradation}. The free-radical mechanism of photo-oxidative degradation is a multistep chain reaction~\cite{yousif2013photodegradation}.
First, incident UV and visible-light radiation excites chromophoric groups within the epoxy resin (e.g. aromatic rings). 
This excitation causes a homolytic dissociation and the formation of radicals, which subsequently abstract hydrogens from other epoxy polymers, leading to further scission and propagation steps~\cite{mailhot2005study}. 
The precise mechanisms after radical formation proposed by~\citet{rivaton1997photo2} proceed in three steps~\cite{rivaton1997photo1, rivaton1997photo2}, illustrated in Fig.~\ref{fig:all-reaction-steps}. 
First, hydrogen abstraction occurs, in which a carbon-hydrogen bond within one of the CH$_3$ groups undergoes homolytic dissociation, leaving the DGEBA molecule with one less hydrogen atom (through the modification of the methyl groups CH$_3$ to CH$_2^\bullet$). 
Then an isomerization reaction occurs in which the constituents of the molecule are rearranged. 
Finally, interactions with UV and visible light cause a scission reaction that breaks DGEBA into an alcohol molecule and quinone methide, producing a strong absorption peak at 340 nm and a second peak in the IR spectrum at 6035 nm (1657 cm$^{-1}$).
As the first step after radical formation in the proposed mechanism of photodegradation of the DGEBA epoxy resin is hydrogen abstraction, computing the activation energy for this process is crucial to understanding the reaction as a whole~\cite{mailhot2005study2}.

\bibliography{main}

\end{document}